\documentclass[aps,prb,twocolumn,superscriptaddress,groupedaddress]{revtex4}

\usepackage{amsmath,amsfonts,bm} %zu 'bm' siehe auguide.ps
\usepackage{graphicx}

\usepackage{color}      %fuer andere Schriftfarben

\newcommand{\ket}[1]{\left. \left| #1 \right. \right\rangle}
\newcommand{\bra}[1]{\left. \left\langle #1 \right. \right|}
\newcommand{\dd}{\mathrm{d}}
\newcommand{\Tr}{\mathrm{Tr}\,}
\newcommand{\imag}{\mathrm{Im}\,}
\newcommand{\fat}[1]{\mathbf{#1}}
\newcommand{\mw}[1]{\left\langle #1 \right \rangle}
\begin{document}

\preprint{exchange/12}

\title{Exchange coupling in transition metal monoxides}

 \author{Guntram  Fischer}
 \affiliation{%
 Institute of Physics, Martin Luther University
 Halle-Wittenberg, Von-Seckendorff-Platz 1, D-06120 Halle, Germany
 }
 \author{Markus D\"ane}
 \affiliation{%
 Max Planck Institute of Microstructure Physics, Weinberg 2,
 D-06120 Halle, Germany
 }
 \affiliation{%
 ORNL, PO BOX 2008 MS6114, Oak Ridge TN 37831-6114, USA
 }
 \author{Arthur Ernst}
 \affiliation{%
 Max Planck Institute of Microstructure Physics, Weinberg 2,
 D-06120 Halle, Germany
 }
 \author{Patrick Bruno}
 \affiliation{%
 Max Planck Institute of Microstructure Physics, Weinberg 2,
 D-06120 Halle, Germany
 }
 \affiliation{%
 European Synchrotron Radiation Facility, BP 220, F-38043 Grenoble Cedex,
 France
 }
 \author{Martin L\"uders}
 \affiliation{%
 Daresbury Laboratory, Daresbury, Warrington, WA4 4AD, UK
 }
 \author{Zdzislawa Szotek}
 \affiliation{%
 Daresbury Laboratory, Daresbury, Warrington, WA4 4AD, UK
 }
 \author{Walter Temmerman}
 \affiliation{%
 Daresbury Laboratory, Daresbury, Warrington, WA4 4AD, UK
 }
 \author{Wolfram Hergert}
 \affiliation{%
 Institute of Physics, Martin Luther University
 Halle-Wittenberg, Von-Seckendorff-Platz 1, D-06120 Halle, Germany
 }

\date{\today}% It is always \today, today,
             %  but any date may be explicitly specified

\begin{abstract}
An \textit{ab initio} study of magnetic exchange interactions in
antiferromagnetic and 
strongly correlated 3$d$ transition metal monoxides is presented.
Their electronic structure is calculated using the local self-interaction 
correction approach, implemented within the Korringa-Kohn-Rostoker band
structure method, which is based on multiple scattering theory. The Heisenberg
exchange constants are 
evaluated with the magnetic force theorem. Based on these the corresponding
N\'eel temperatures $T_N$ and spin wave dispersions are calculated.
The N\'eel temperatures are obtained using mean field approximation,
random phase approximation and Monte Carlo simulations.
The pressure dependence of $T_N$ is investigated using exchange constants
calculated for different lattice constants. All the calculated
results are compared to experimental data.
\end{abstract}

\maketitle

\section{Introduction \label{sec:introduction}}
In the last years there has been a general strong interest in finding
materials with specific or even parametrisable magnetic properties. Such
materials could be useful in the field of spintronics. A lot of
the promising candidates are strongly correlated electronic systems
which in many ways are still a challenge to be properly described theoretically
regarding their electronic ground state properties. 
On the
other hand, for reliable predictions about magnetic properties of materials it
is essential to have theories describing the magnetism adequately by
quantitative and qualitative means. One of these theories is the Heisenberg
theory of magnetism, which we shall apply in the present paper. Its central
quantities are the Heisenberg exchange constants $J_{ij}$, which are of general
fundamental interest. In particular they provide information about the
magnetic periodicity (via their Fourier transform), the spin-wave dispersion,
magnetic critical temperatures and also allow predictions on structural
effects caused by magnetism\cite{PhysRevLett.87.255501,l652}.\\
%%%%%%%%%%%%%%%%%%%%%%%%%%%%%%%%%%%%%%%%%%%%%%%%%%%%%%%%%%%%%%%%%%%%%%%%%%%%%%%
In this paper we concentrate on the study of the magnetic exchange
interactions of
transition metal monoxides (TMOs), specifically MnO, FeO, CoO and NiO. They
are charge-transfer insulators, well known for strong correlation effects
associated with the TM 3$d$-electrons. Originating from the Anderson-type
superexchange, their equlibrium magnetic structures
are of the antiferromagnetic II (AFII) order, characterized by planes of
opposite magnetization which are stacked in (111)-direction. Recently, a Mott
transition has been observed in MnO at high 
pressure of about 105 GPa, in resistivity\cite{p04} and X-ray spectroscopy
measurements,\cite{ymk05,mrb07} which stimulated new theoretical studies in
this high pressure region. \cite{kasinathan:195110,wl08} There already exists
a large body of neutron 
scattering measurements of magnetic structures and magnetic excitations in
transition metal monoxides. However, the development of new experimental
techniques such as neutron powder diffraction\cite{gtc05,gtd06,gdt07} and
polarized neutron reflectivity\cite{o08} has renewed interest in studying TMOs
as antiferromagnetic benchmark materials. Modern neutron spectrometers operate
with such a high efficiency that also high angle diffraction experiments can
be performed to unravel complex magnetic order e. g. in thin
films\cite{zib00}.\\ 
%%%%%%%%%%%%%%%%%%%%%%%%%%%%%%%%%%%%%%%%%%%%%%%%%%%%%%%%%%%%%%%%%%%%%%%%%%%%%%%
From the theory point of view conventional methods such as the local spin
density approximation (LSDA) to density functional theory (DFT), treating
electron correlations at the level of the homogeneous electron gas, fail to
provide an adequate description of the electronic structure of these
oxides. Over the years a number of approaches has been developed, aiming at
improvements to the LSDA treatment of electron correlations, and applied to
TMOs with varying degrees of success. Among them are: the LSDA+U
method\cite{PhysRevB.62.16392,PhysRevB.69.075413}, GGA+U\cite{zhh06},
self-interaction corrected
(SIC)-LSDA\cite{Temmerman98,tss00,sg90,stw93,Diemo_NiO-Paper,TMO-SIC-paper},
hybrid functionals\cite{franchini:045132,f04}, and finally
dynamical mean field theory\cite{kas07}. In general, they have
improved lattice constants, band gaps and magnetic properties, some of
them have also obtained good agreement with spectroscopies. \\ 
%%%%%%%%%%%%%%%%%%%%%%%%%%%%%%%%%%%%%%%%%%%%%%%%%%%%%%%%%%%%%%%%%%%%%%%%%%%%%%%
In the present paper we shall use the so-called local 
self-interaction correction\cite{pz81} (LSIC) scheme for the calculation of
the electronic ground states of the TMOs. As the aim is the investigation of
magnetic interactions we combine the LSIC scheme with the magnetic force
theorem (MFT)\cite{lka87} in order to obtain the Heisenberg exchange
parameters $J_{ij}$.\\
The LSIC scheme\cite{LED05} is based on the implementation
of the SIC-LSDA formalism\cite{pz81,Temmerman98,tss00} within multiple
scattering theory in the framework of the Korringa-Kohn-Rostoker (KKR)
band structure method. It was first applied to \textit{f}- electron systems
\cite{LED05,hde07}, but recently also to TMOs\cite{TMO-SIC-paper}. Within the
KKR method one calculates the Green function of the investigated system.
This Green function is then straightforwardly used in the application of the
MFT. It is for the first time that this combined approach is applied for
calculating exchange constants of the transition metal monoxides. The results
of that are compared to the exchange constants extracted from the total energy
differences for a number of magnetic structures and mapping them onto a
Heisenberg Hamiltonian. Most of the earlier applications of the latter
approach have been based on the assumption that only the first two exchange
interaction constants are nonzero. Although the present combined
approach also relies on the mapping onto a Heisenberg Hamiltonian the
assumptions regarding the number of non-zero exchange constants are not
needed, which is advantageous to systems with reduced symmetry such as thin
films and layered structures (where the justification for such an assumption
is not clear from the very beginning).\\
Having calculated the $J_{ij}$ for the ground states of the TMOs we
also calculate and discuss them as a function of external pressure for moderate
values of the latter. This is mainly inspired by the recent high 
pressure measurements of TMOs\cite{p04,ymk05,mrb07}. \\
Based on the calculated magnetic exchange
interactions the transition temperatures can be obtained. In this paper it is
done in three different ways, namely by applying mean-field approximation
(MFA), random phase approximation (RPA), and using classical Monte Carlo (MC)
simulations. The respective results are then compared to those obtained from
the disordered local moments (DLM) method\cite{DLM-paper}, which does not
involve mapping onto a Heisenberg Hamiltonian, but is based on the same ground
state electronic structure calculations as the present
paper\cite{TMO-SIC-paper}.\\
The last subject we focus on are magnetic excitations. 
On one hand, with given $J_{ij}$, one can calculate the magnon
spectrum of any material. On the other hand, measuring the latter
experimentally is a direct method to examine its exchange constants. Thus,
comparing calculated to experimental spin wave dispersions provides a
straightforward tool for determining the accuracy of 
the calculated $J_{ij}$.\\
%%%%%%%%%%%%%%%%%%%%%%%%%%%%%%%%%%%%%%%%%%%%%%%%%%%%%%%%%%%%%%%%%%%%%%%%%%%%%%%
The present paper is organized as follows: In section \ref{sec:theory} the
theoretical approaches for the calculation of electronic structure, exchange
interactions and N\'eel temperatures are presented. The computational details
are described in section \ref{sec:comp-deta}. Section \ref{sec:results}
contains the results and discussion. The exchange parameters and the N\'eel
temperatures are presented for theoretical equilibrium lattice constants and
as a function of lattice constants and pressure, respectively. Finally, the
calculated  magnon spectra of the TMOs are discussed in reference to
experiments. The paper is concluded in section \ref{sec:conclusion}. 

\section{Theory \label{sec:theory}}
\subsection{Electronic Structure}

For the electronic structure calculations of TMOs we use a multiple
scattering theory-based implementation of the SIC-LSDA method\cite{pz81}, whose
total energy functional is 
%%%%%%%%%%%%%%%%%%%%%%%%%%%%%%
\begin{eqnarray}
  E^{\mathrm{SIC-LSDA}}[\{n_{\alpha \sigma}\}] &=&
    \tilde{E}^{\mathrm{LSDA}}\left[ n_{\uparrow},n_{\downarrow} \right]-\nonumber \\
    \label{eq:E-SIC-LSDA} &&
    \sum_{\alpha \sigma} \left(
    E_{\mathrm{H}}\left[n_{\alpha}\right] +
%   E_{\mathrm{H}}\left[n_{\alpha \sigma}\right] +
    E^{\mathrm{LSDA}}_{\mathrm{xc}}\left[n_{\alpha \sigma},0 \right] \right)
  \; ,\quad ~
\end{eqnarray}
%%%%%%%%%%%%%%%%%%%%%%%%%%%%%%
with the LSDA energy functional in units of Rydberg given by
%%%%%%%%%%%%%%%%%%%%%%%%%%%%%%
\begin{eqnarray}
    \tilde{E}^{\mathrm{LSDA}}\left[ n_{\uparrow},n_{\downarrow} \right] &=&
    \sum_{\alpha \sigma} \bra{\phi_{\alpha \sigma}}-\nabla^2
    \ket{\phi_{\alpha \sigma}} + \quad \quad \quad
    \nonumber \\ \label{eq:E-LSDA}
    &&  E_{\mathrm{ext}} + E_{\mathrm{H}}[n] +
    E^{\mathrm{LSDA}}_{\mathrm{xc}}\left[ n_{\uparrow},n_{\downarrow} \right]
    \; .\quad \quad
\end{eqnarray}
%%%%%%%%%%%%%%%%%%%%%%%%%%%%%%
Here $\phi_{\alpha \sigma}$ is a Kohn-Sham orbital, $\alpha \sigma$
a multi-index labelling the orbitals and spin ($\uparrow$ or
$\downarrow$), respectively, $n_{\alpha \sigma}=|\phi_{\alpha
\sigma}|^2$, $n_{\sigma}=\sum_{\alpha}n_{\alpha \sigma}$ and $n =
n_{\uparrow}+n_{\downarrow}$. $\tilde{E}^{\mathrm{LSDA}}$ differs from
$E^{\mathrm{LSDA}}$ since the kinetic energy is evaluated with respect to the
orbitals minimizing the SIC-functional. The summations run over
all the occupied orbitals, $ E_{\mathrm{ext}}$ denotes the external energy
functional due to ion cores, $E_{\mathrm{H}}$ is the Hartree 
energy and $E^{\mathrm{LSDA}}_{\mathrm{xc}}$ is the LSDA
exchange-correlation energy functional.
%%%%%%%%%%%%%%%%%%%%%%%%%%%%%%
The second term in Eq.~(\ref{eq:E-SIC-LSDA}) is the so-called 
self-interaction correction\cite{pz81} for all the occupied orbitals $\alpha$.
It restores the property 
%%%%%%%%%%%%%%%%%%%%%%%%%%%%%%
\begin{equation}
  \label{eq:E_H-E_xc-cancellation}
  E_{\mathrm{H}}[n_{\alpha}] + E^{\mathrm{exact}}_{\mathrm{xc}} [n_{\alpha,
    \sigma},\, 0] = 0 \; ,
\end{equation}
%%%%%%%%%%%%%%%%%%%%%%%%%%%%%%
that the exact DFT exchange-correlation functional has, namely
that for any single orbital density the Hartree term should be cancelled by the
corresponding exchange-correlation term. The cost paid for restoring
the above property is the orbital dependence of the SIC-LSDA energy functional
(Eq.~(\ref{eq:E-SIC-LSDA})). The correction is only substantial for localized
orbital states, but vanishes for itinerant states. In the limit of all itinerant
states the SIC-LSDA total energy functional is identically equal to the LSDA
functional.\\
%%%%%%%%%%%%%%%%%%%%%%%%%%%%%%
The main idea behind the ``local'' implementation of the SIC-LSDA
formalism (LSIC) is that within multiple scattering theory, in the framework
of the KKR method, one works with the scattering phase shifts, describing
scattering properties of single atoms in a solid. Among them only the
resonant phase shifts are relevant, as they refer to localized states. 
Thus the self-interaction 
correction is associated with the on-site scattering potentials and
leads to modified resonant scattering phase shifts. In particular, they become
stronger localized. Details of the LSIC implementation are discussed in
Ref.~\onlinecite{LED05}.

\subsection{Magnetic Interactions}

The Heisenberg theory of magnetism assumes that it is possible to map magnetic
interactions in a material onto localized spin moments, which in a classical
picture can be represented by a vector. The resulting classical
Hamiltonian,
\begin{equation}
  \label{eq:Ham-Rusz}
  H = - \sum_{ij} J_{ij} \fat{e}_{i} \cdot  \fat{e}_{j} \; ,
\end{equation}
contains only the unit vectors $\fat{e}_{i(j)}$ of
the spin moments and the exchange parameters $J_{ij}$ describing the 
interactions between them\cite{Hamiltonian_note}. Here $i$ and $j$ index the
sites. \\
It should be mentioned here that the Hamiltonian (\ref{eq:Ham-Rusz}) can be
extended to include additional effects like magnetocrystalline anisotropy or
tetragonal or rhombohedral distortions of the lattice. The latter reflect
magnetoelastic effects which result in two different values for the nearest
neighbour (NN) exchange parameters, depending on a parallel or antiparallel
alignment of the moments. Such effects are usually present in
experiments. Thus, special care is required when comparing theoretical and
experimental results.\\
Our method of choice for the calculation of the
exchange parameters $J_{ij}$ makes use of the magnetic force theorem, but
invokes also mapping onto a Heisenberg Hamiltonian. For comparison, we also
apply the most commonly used approach which relies on the calculation of
total energy differences between different magnetic configurations and
mapping them onto a classical Heisenberg Hamiltonian.

\subsubsection{Magnetic Force Theorem Approach}

The idea behind the magnetic force theorem\cite{lka87} is to consider
infinitesimally small rotations of classical spins at two different lattice
sites. These give rise to energy changes that are mapped onto the
classical Heisenberg Hamiltonian via multiple scattering theory. This approach
is based on the assumption that the potentials are unchanged by the
rotations. The advantage of the MFT method is that the exchange integrals can
be calculated directly in the relevant magnetic structure. The result for 
the exchange parameter $J_{ij}$ of the two magnetic moments at
sites $i$ and $j$ can be written as
\begin{equation}
  \label{eq:MFT}
  J_{ij} = \frac{1}{8 \pi} \int^{\epsilon_F} \dd\epsilon\, \imag \Tr_L
  \left( \Delta_i \hat{\tau}_{\uparrow}^{ij} \Delta_j
    \hat{\tau}_{\downarrow}^{ji} +
    \Delta_i \hat{\tau}_{\downarrow}^{ij} \Delta_j
    \hat{\tau}_{\uparrow}^{ji} \right) \; ,
\end{equation}
where $\hat{\tau}^{ij}$ is the scattering path operator between
sites $i$ and $j$ and $\Delta_i = \hat{t}_{i \uparrow}^{-1} -
\hat{t}_{i \downarrow}^{-1}$, with $\hat{t}_i$ being a single scattering
operator for the atom at site $i$. If not stated otherwise, all the results
discussed later would have been obtained with the exchange parameters
calculated using Eq.~(\ref{eq:MFT}).

\subsubsection{Energy Differences Approach}

\begin{figure}[htbp]
  \centering
  \includegraphics[width=8cm]{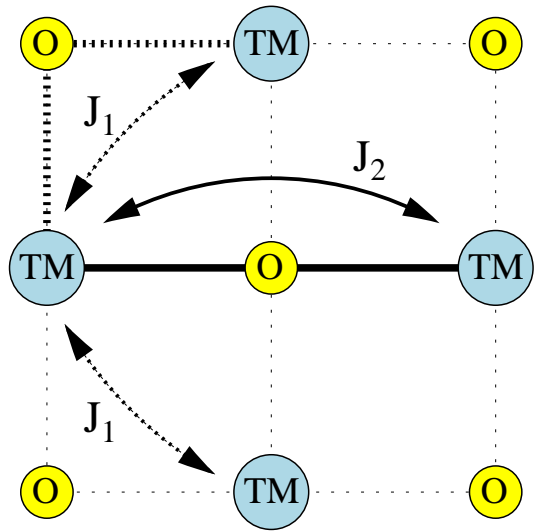}
  \caption{\label{fig:J1J2diagramm}
    (color online) Schematic representation of the magnetic interactions in a
    (100) plane of the rocksalt structure of TMOs. The TM ions (blue) interact
    via $J_1$ (dotted arrows) with their nearest and via $J_2$ (solid arrow)
    with their next nearest neighbours. In Anderson's super-exchange picture
    the indirect exchange is mediated by O ions (yellow circles), resulting in
    a $90^{\circ}$- and a $180^{\circ}$- exchange interaction for $J_1$ and $J_2$,
    respectively. Note that $J_1$ also contains contributions
    from direct overlap.    
  }
\end{figure}

In this approach the total energies of the TMOs in the ferromagnetic (FM) and
antiferromagnetic I and II (AFI and AFII) configurations are taken
into account. The AFI structure is characterized by oppositely magnetized
planes which are stacked in (100)-direction. Suppose that magnetic
interactions operate only between TM 
atoms --- an assumption which is to be discussed later --- the mapping onto the
Heisenberg Hamiltonian yields
\begin{equation}
  \label{eq:J1}
  J_1=\frac{1}{16} \left( E_{\mathrm{AFI}} - E_{\mathrm{FM}} \right)
\end{equation}
and
\begin{equation}
  \label{eq:J2}
  J_2=\frac{1}{48} \left( 4 E_{\mathrm{AFII}} - 3 E_{\mathrm{AFI}} -
    E_{\mathrm{FM}} \right) \;,
\end{equation}
where $J_1$ describes the interaction between the NN and $J_2$ that of the
next nearest neighbours (see Fig. \ref{fig:J1J2diagramm}). This
mapping also assumes that the interaction between NN is independent of the
sublattice the TM atoms are located on. 
Of course the choice of the three above mentioned structures restricts one
to the determination of $J_{1}$ and $J_{2}$ only. Using more magnetic
structures and hence calculating more exchange parameters is in principle
possible. However, due to the nature of the present exchange mechanism, the
super exchange, this has usually not been done for the TMOs. Although with this
method we also restrict exclusively to $J_{1}$ and
$J_{2}$, it is hoped that the comparison with the MFT method will shed some
light on the validity of the underlying assumptions for the TMOs.

\subsection{N\'eel Temperatures}

Having calculated the exchange parameters for an
antiferromagnet one is able to calculate the N\'eel temperatures,
$T_{\mathrm{N}}$. Several 
different approximations can be used, in particular the mean field
approximation, the random phase approximation and classical Monte Carlo
simulations. 

\subsubsection{Random Phase Approximation}

In the RPA one solves the equation of motion for the Green function of
the spin operators. Following the approach of Rusz \textit{et
  al.}\cite{rtd05}, one ends up with a semi-classical formula for the
average spin polarization $\mw{e_{A}^z}(T)$ of a sublattice (e.g. $A$), as a
function of temperature $T$,
\begin{equation}
  \label{eq:<s>(T)}
  \mw{e_{A}^z}(T) = \mathcal{L} \left( \frac{2}{k_B T} \left[
      \frac{1}{\Omega}
      \int \dd \fat{q} \left[ \mathbf{N}^{-1} (\fat{q}) \right]_{AA}
    \right]^{-1} \right) \; .
\end{equation}
Here $\mathcal{L}(x)$ is the Langevin function, $\mathcal{L}(x)=\coth(x)-1/x$,
$\Omega$ is the volume of the first Brillouin zone and $k_B$ denotes the
Boltzmann constant. The matrix elements of $\mathbf{N}(\fat{q})$ are defined as
\begin{equation}
  \label{eq:N-Matrix}
  N_{AB}(\fat{q}) = \delta_{AB} \sum_C J_{AC}(0)
    \mw{e_{C}^z} \;- \mw{e_{A}^z} J_{AB}(\fat{q}) ,
\end{equation}
with the Fourier transforms of the exchange parameters given by
\begin{equation}
  \label{eq:J_FT}
  J_{AB}(\fat{q}) = \frac{1}{\nu} \sum_{j,k} J_{jk} e^{ i\fat{q}\cdot
    (\fat{R}_{j}-\fat{R}_{k})} \gamma_j^A \gamma_k^B \;, 
\end{equation}
where $\nu$ denotes the number of interacting magnetic sites and $\gamma_j^A$
equals one if site $j$ is on the magnetic sublattice $A$ and zero otherwise.
Eq.~(\ref{eq:<s>(T)}) has to be
solved self-consistently since the unknown quantity appears on its left
and implicitly also on the right, via $\mathbf{N}(\fat{q})$. 
The N\'eel temperature is equal to the highest value of $T$ 
at which $\mw{e_{A}^z}(T)$ becomes different from zero.

\subsubsection{Mean-Field Approximation}

To obtain the MFA estimate of the N\'eel temperature, a matrix
$\mathbf{\Theta}$ with elements
\begin{equation}
  \label{eq:Theta_AB}
  \Theta_{AB}= \frac{2}{3 k_B} J_{AB}(0) \; 
\end{equation}
is constructed \cite{ssb04,Anderson_SolidStatePhysics14}, where $J_{AB}(0)$
stands for the Fourier 
transform of the exchange para\-meters, defined via Eq.~(\ref{eq:J_FT}), at
$\fat{q}=0$. The largest eigenvalue of 
$\mathbf{\Theta}$ yields the N\'eel temperature. 
If for any TMO in the AFII structure only the nearest and
next-nearest neighbour interactions are considered ($J_1$ and $J_2$,
respectively), then the largest eigenvalue yields the
well-known relation, $k_B T_\mathrm{N} = 4 J_2$, indicating that the nearest
neighbour interaction $J_1$ does not have any influence on $T_\mathrm{N}$. Since
fluctuations are completely neglected in MFA the resulting N\'eel
temperatures are commonly overestimated. 

\subsubsection{Monte Carlo Simulations}

We give a rather brief summary of the
method of MC simulations as they are performed in this paper. For a
deeper and complete understanding we refer the reader to the book by Landau
and Binder\cite{MC-Buch-neu}.\\
To estimate $T_N$ via MC simulations a lattice representing the structure
of the investigated system is constructed. The magnetic
moment at lattice site $i$ interacts with its neighbours $j$ via the
$J_{ij}$. During a MC run one picks a lattice site $j$ with
the magnetic moment vector $\fat{e}_j$, creates a new random direction
$\fat{e}'_j$ and decides by looking at the energy of the system whether
$\fat{e}'_j$ is accepted or $\fat{e}_j$ is kept. Performing this
procedure $N$-times on a lattice of $N$ sites is defined as one MC step.\\
Starting from a certain initial configuration the system is
brought into thermal equilibrium for a fixed 
temperature. After this, ``measurements'' and thermodynamical 
averaging of the observables of interest are performed. 
Since it is impossible during a simulation run to go through all possible
configurations of the system, which would be formally necessary for averaging,
one must ensure that the configurational subspace that one is restricted to is
of physical significance. This is done by performing the so-called importance
sampling. It is applied when one has to decide between the old and new
magnetic moment vectors $\fat{e}'_j$ and $\fat{e}_j$ described above. There
exist several methods to do this, in the present paper the Metropolis
algorithm\cite{Metropolis} is used.\\
One  must be aware of the fact that the finite size of the
lattice, despite being periodic in all 3 dimensions, leads to a systematic
error in the determination of the critical temperature. This so-called
finite-size effect, however, becomes smaller with increasing lattice size.
It can therefore be eliminated by extrapolating the critical temperatures for
different lattice sizes.\\
For a magnetic system as in the present case it is
straightforward to measure two quantities. One is the staggered
magnetization\cite{PhysRevB.38.6868} $\fat{m}_s$ being some sort of an average
of the absolute values of magnetization of the sublattices, 
\begin{equation}\label{M_st}
  \fat{m}_s = \frac{1}{N} \sum_{j=1}^N \fat{e}_j \,e^{i \fat{Q}\cdot
    \fat{r}_j}\; .
\end{equation} 
Here, $j$ labels the lattice sites being $N$ in total, $\fat{e}_j$ is the
unit vector of the magnetic moment at lattice site $j$ and $\fat{Q}$ is the
normal vector of the planes of equal magnetization, in the AFII structure
being $(1,\, 1,\, 1)$ for example. As a note, $\fat{m}_s$ is used instead of
the total magnetization since the latter is equal to zero in
antiferromagnets. The other quantity measured is the inner, i.e.\
magnetic, energy of the system $E$, which is given by Eq.~(\ref{eq:Ham-Rusz}). 
This whole procedure of relaxing into thermal equilibrium and thermodynamical
averaging is repeated for different temperatures. In principle one can
determine $T_N$ from the slope of the temperature dependence of $\fat{m}_s(T)$
and $E(T)$. However, there are quantities that show the critical temperatures
more clearly. These are the magnetic susceptibility derived from $\fat{m}_s$,
the 
specific heat derived from $E$, and the 4th-order cumulant\cite{Binder-Kum}
$U_4$. The first two have a singularity at $T=T_N$, the last one has the
property that the curves $U_4(T)$ calculated for different lattice sizes have
a crossing point at $T=T_N$. 

\subsubsection{Quantum Effects}

As already mentioned, the three approaches described above are based on
mapping of the single magnetic moment interactions onto a classical Heisenberg
Hamiltonian. These moments, however, are quantum objects and this should in
some way be accounted for in the calculations. Wan, Yin and
Savrasov\cite{wys06} and Harrison \cite{h07} did this by replacing the
classical $S^2$ in the 
Heisenberg Hamiltonian with the quantum mechanical expectation value
$S(S+1)$, when calculating magnetic properties. Since in
Eq.~(\ref{eq:Ham-Rusz}) $S^2$ is included in the 
$J_{ij}$\cite{Hamiltonian_note}, then to be consistent, one has to divide
again by $S^2$. This gives rise to a factor $(S+1)/S$ for the energy and,
eventually, also for the N\'eel temperature\cite{QuantumScaling_note}. This
factor is, however, not needed when the  N\'eel temperature is
obtained based on the DLM method since it does not explicitely use the
$J_{ij}$. 

\subsection{Magnon Spectra}

With the exchange interactions determined one can also calculate the magnon
spectra $E(\fat{q})$. They are of special interest since they provide the
standard method for determining the exchange parameters experimentally. The
latter would be done by fitting a Hamiltonian containing the $J_{ij}$ as
fitting parameters to a measured spin wave dispersion. As already mentioned in
section \ref{sec:theory}, such Hamiltonians usually contain more terms than the
one given in Eq.~(\ref{eq:Ham-Rusz}), which is why comparisons between
different $J_{ij}$ results must be done carefully.\\
Considering multiple sublattices one can define magnon spectra as the
eigenvalues of the matrix $N(\fat{q})$ given by
Eq.~(\ref{eq:N-Matrix}). Assuming two magnetic 
sublattices, with the same absolute magnetization, and considering only the
nearest and next-nearest neighbour interactions, the spectra are given
by\cite{st98,Hamiltonian_note}
\begin{equation}
  \label{eq:Terakuraformel}
  E(\fat{q})= \frac{1}{\mu} \sqrt{\left(J_{++}(\fat{q}) - H_0 \right)^2 -
              J_{+-}^2(\fat{q})} \;.
\end{equation}
Here $\mu$ is the magnetic moment of the two sublattices in units of $\mu_B$,
$J_{++}(\fat{q})$ 
($J_{+-}(\fat{q})$) are the Fourier transforms of the intra- (inter-)sublattice
exchange parameters expressed respectively as
\begin{eqnarray}
  \label{eq:Jpp}
  \lefteqn{J_{++}(\fat{q}) = 2 J_1 \times} \\ 
 &&\times \left( \cos \pi a(q_y + q_z) + \cos \pi a(q_x + q_y)
   % \right. \\
%  && \left. 
   + \cos \pi a(q_x + q_z) \right) \nonumber
\end{eqnarray}
and
\begin{eqnarray} 
  \label{eq:Jpm}
  \lefteqn{J_{+-}(\fat{q}) = 2 J_1 \times}\\
 &&\times \left( \cos \pi a (q_y - q_z) + \cos \pi a (q_y - q_x) +
 % \right.\\
%  && \left.
      \cos \pi a (q_x - q_z) \right)\nonumber \\ 
 &&{} + 2 J_2 \left(\cos 2\pi a\, q_x + \cos 2 \pi a\, q_y + \cos 2 \pi a\, q_z
 \right)\, , \nonumber 
\end{eqnarray}
and $H_0 = J_{++}(0)-J_{+-}(0)= - 6 J_2$. In Eqs.~(\ref{eq:Jpp}) and
(\ref{eq:Jpm}) the vector $\fat{q}$ and accordingly its components are needed
in units of $2 \pi /a$ with $a$ being the lattice constant of the TMO
considered.

\section{Computational details}
\label{sec:comp-deta}

The transition metal monoxides crystallize in the rocksalt structure (B1,
Fm$\overline{3}$m, space group 225). At low temperatures they show 
small lattice distortions ($< 2 \%$). However, these distortions are not
considered in the present calculations. The crystal potentials for the
ground state calculations are constructed in the atomic sphere
approximation (ASA). The ASA radii for the TM and oxygen atoms
are chosen as $0.2895\, a$, with $a$ being the lattice constant of
a given TMO. To reduce the ASA overlap while keeping a
good space filling, empty spheres are used with the ASA
radii equal to $0.1774\, a$. The ratios of the respective ASA radii are
kept constant across the TMO series.\\
For the electronic structure calculations
the complex energy contour has 24 Gaussian quadrature points, and for the
Brillouin zone (BZ) integrations a 14x14x14 k-points mesh is
constructed. For the calculation of the magnetic interactions, using the MFT, 60
energy points on a Gaussian mesh in the complex plane are
chosen. Convergence of the $J_{ij}$ with respect to the number of k-points is
achieved with a 20x20x20 k-points mesh per energy point for the first 50 
of them, and a 60x60x60 k-mesh for the last 10 energy points, lying close to
the Fermi energy.\\
For the MC simulations an fcc-lattice representing the transition metal atoms
in the TMO crystal is constructed. To avoid finite-size effects, the size of
the lattice is varied from 40x40x40 to 60x60x60 elementary fcc cells.
To use all observables described in the MC part of Section
\ref{sec:theory} one has to restrict the simulations to a relatively small
number of MC steps. This is necessary in order to prevent the system from
changing the orientation of the ferromagnetic sublattices, for example from
($111$) to ($\bar{1}11$), which are degenerate in energy. Thus, starting
from the AFII state, the system is assumed to have reached
thermal equilibrium after 5,000 MC steps, and for averaging 10,000 MC steps
are performed. If one does so all
observables yield the same result for the N\'eel temperatures for each TMO,
respectively. To ensure a 
thorough exploration of phase space, simulations with up to 100,000 MC
steps for averaging have also been performed. In this case the specific heat,
not affected by reorientations of the magnetic sublattices, has reassuringly
indicated the magnetic phase transitions to occur at the same temperatures as 
in the short simulations.

\section{Results and Discussion \label{sec:results}}
\subsection{Exchange parameters}

The present calculations of the exchange parameters of TMOs use the ground
state electronic structure properties of these materials as input. The
latter are  obtained self-consistently with the LSIC method, explained in
detail in Ref.~\onlinecite{TMO-SIC-paper}. In particular, as seen in
Eq.~(\ref{eq:MFT}), for the MFT approach the relevant quantities
are the scattering properties evaluated at the equilibrium lattice constant of
the ground state, AFII, magnetic structure. For the approach based on the
energy differences only the total energies of the FM, AFI and AFII structures,
evaluated at the theoretical equilibrium lattice constants of the AFII
configuration, are of relevance. 
\begin{table}[htbp]
  \caption{\label{tab:a0}
    The calculated (calc.) equilibrium lattice constants and spin magnetic
    moments, per TM-atom, for all the studied TMOs in AFII structure. The oxygen
    atoms are not spin polarized in the AFII environment and the induced
    moments on the empty spheres are very small. Consequently the calculated
    spin magnetic moments of TMOs are practically equal to those of their
    TM-atoms. The experimental (exp.) values of the magnetic moments contain
    not only the spin but also orbital contribution.
  }
  \begin{tabular}{cc*{2}{r}*{2}{r}} \hline
    &\hspace{1em} &
      \multicolumn{2}{c}{$a_0$  [\AA]} &
      \multicolumn{2}{c}{$\mu$  [$\mu_B$]} \\ \cline{3-6}
    \raisebox{2.1ex}[1.7ex][0ex]{~TMO~} &
    & \multicolumn{1}{r}{calc.~} &\multicolumn{1}{c}{~exp.} &
      \multicolumn{1}{r}{~calc.~} & \multicolumn{1}{c}{~exp.} \\ \hline
    MnO && 4.49 & 4.44 \cite{landolt-b.III.27g}
    & 4.63 & ~4.54 \cite{PhysRevB.67.184420}\\
    FeO && 4.39 & ~4.33 \cite{fgs96}
    & 3.68 & 3.32 \cite{PhysRev.110.1333} \\
    CoO && 4.31 & 4.26 \cite{landolt-b.III.27g}
    & 2.69 & 2.40 \cite{jr02} \\
    NiO && 4.24 & 4.17 \cite{landolt-b.III.27g}& 1.68
    & 1.90 \cite{PhysRevB.27.6964} \\ \hline
  \end{tabular}
\end{table}

\begin{figure}[htbp]
  \centering
  \includegraphics[width=8cm]{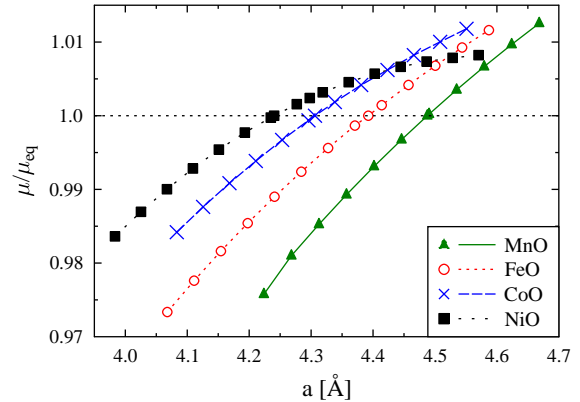}
  \caption{\label{fig:mws_all} (color online)
    The dependence of the calculated spin magnetic 
    moments on lattice constants. The spin magnetic moments have been divided
    by their respective equlibirium values given in Table
    \ref{tab:a0}. The crossing points of each curve with the horizontal
    dotted line at $\mu/\mu_{eq}=1$ mark the equilibrium lattice constant of
    each TMO.}
\end{figure}

From Table \ref{tab:a0} we can see that the LSIC method, treating 
localized and itinerant electrons on equal footing, reproduces well the
equilibrium 
lattice constants and also the corresponding spin magnetic moments in the AFII
structure. %as 
The overall agreement with the experimental values is reasonable
for both quantities. Note, however, that the experimental magnetic moments
listed in the table include both the spin- and orbital-contributions, which
are substantial for FeO and CoO, and non-negligible even for NiO. Regarding
the calculated spin magnetic moments, they are effectively equal to the spin
moments of the TM atoms, as the oxygen atoms are not polarized in the AFII
environment, and the induced spin moments on the empty spheres are very
small. In addition, as seen in Fig.~\ref{fig:mws_all}, the spin magnetic
moments show considerable dependence on the lattice constants, indicating that
a similar behaviour may also be expected for the calculated exchange constants.
This dependence of the calculated spin magnetic moments on the lattice constant
was also observed in previous studies\cite{fss99,kasinathan:195110,zhh06} and
its trend agrees with the Stoner model, stating that magnetic moments
eventually collapse at very high pressures.\\
Using the above ground state properties in the MFT approach, we have
calculated the $J_{ij}$ exchange constants for the first 11 neighbour
shells. As expected only the first two of them, $J_{1}$ and $J_{2}$, are of
relevance as those corresponding to the higher shells are less than
$0.1\,\mathrm{meV}$ in magnitude. This agrees well
with the idea of super-exchange \cite{Anderson_SolidStatePhysics14} and
can easily be explained with it. Consequently, and for the purpose of
comparison with the results of the energy difference approach, in Table
\ref{tab:J_series} we display only the $J_{1}$ and $J_{2}$ quantities, as well
as the 
experimental results. Our results also provide justification for one of
the assumptions underlying Eqs.~(\ref{eq:J1}) and (\ref{eq:J2}), that the
interactions between NN atoms are not dependent on the sublattices of the atoms.
\begin{table*}[htbp]
  \caption{\label{tab:J_series}
    The exchange parameters, $J_i$, in meV, with $i$ being the shell index,
    for the first two shells of all the studied TMOs and both the MFT and 
    $\Delta E$, energy difference, methods. The experimental (exp.) values are
    given in the leftmost column, respectively, according to the
    Hamiltonian\cite{Hamiltonian_note} in Eq.~(\ref{eq:Ham-Rusz}).
    For $i>2$, the absolute magnitudes of the $J_i$'s for all the TMOs have
    been less than $0.1\,\mathrm{meV}$. For both the MFT and $\Delta$E
    approaches the 
    calculated equilibrium lattice constants from Table \ref{tab:a0} have been
    used. The results of Harrison\cite{h07}, obtained using a tight-binding (TB)
    formalism for the complete series of TMOs are listed here for a direct
    comparison. The other previous results, obtained for selected monoxides,
    are listed in Table \ref{tab:J_single}.
  }
  \begin{tabular}{cccccrccccc}
  \hline
  &~~&\multicolumn{4}{c}{$J_{1}$}&~~&\multicolumn{4}{c}{$J_{2}$}\\
  \cline{8-11} \cline{3-6}  
  \raisebox{2.1ex}[1.7ex][0ex]{~TMO~} && exp.&~ MFT~ &~ $\Delta$E~ \,&
  \multicolumn{1}{c}{~~TB\cite{h07}}&& 
    \, exp.&~ MFT~ &~ $\Delta$E~ \,& \multicolumn{1}{c}{~~TB\cite{h07}}\\\hline 
  ~MnO&& -2.06,\,-2.64\cite{p74}&-0.91 & 0.68 & -4.41~ &&
        -2.79\cite{p74}        &-1.99 &-1.65 & -1.09~\\
  ~FeO&& 1.04,\,1.84\cite{khc78}&-0.65 & 0.48 & -2.99~ &&
       -3.24\cite{khc78}       &-3.17 &-3.50 & -1.56~\\
  ~CoO&& 0.70\cite{ti06},\,-1.07\cite{PhysRevB.6.4294}  &
       -0.32 & 0.53 & -1.83~ &&
       -6.30\cite{ti06},\,-5.31\cite{PhysRevB.6.4294}  &
       -4.84 &-4.40 & -1.64~\\ 
  ~NiO&&-0.69\cite{PhysRevB.7.5000},\,0.69\cite{Hutchings_Samuelson} &
       ~0.15 & 1.42& -1.44~ &&
      -8.66\cite{PhysRevB.7.5000},\,-9.51\cite{Hutchings_Samuelson}&
      -6.92 &-6.95 & -1.88~\\ \hline
  \end{tabular}
\end{table*}\\

\begin{figure}[htbp]
  \centering
  \includegraphics[width=8cm]{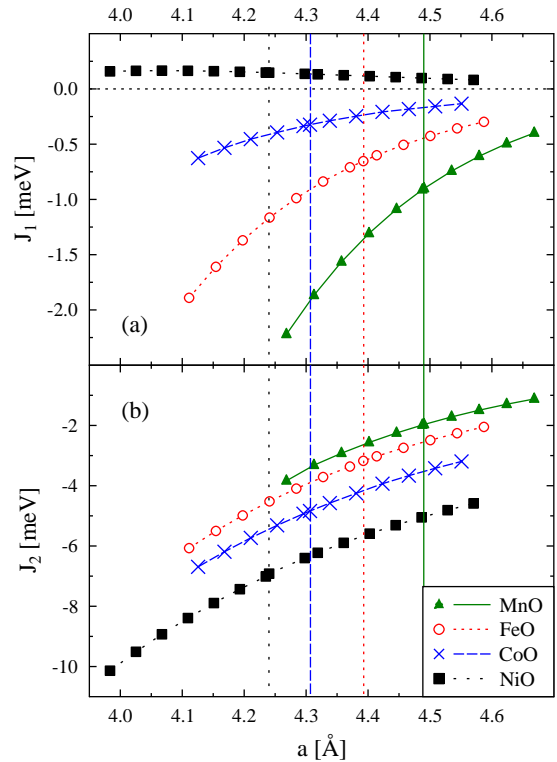}
  \caption{\label{fig:J_all} (color online)
    (a) The $J_1$- and (b) $J_2$-dependency on the
    lattice constant $a$ for all TMOs, calculated using MFT. The vertical
    lines mark the calculated equilibrium lattice constants from Table
    \ref{tab:a0}.} 
\end{figure}

\begin{table*}[htbp]
  \caption{\label{tab:J_single}
    Summary of the \textit{first principles} results for $J_1$ and $J_2$ in
    MnO and NiO, based on the Hamiltonian\cite{Hamiltonian_note} in
    Eq.~(\ref{eq:Ham-Rusz}), from the present and previous theoretical works
    for comparison. Only those close to 
    experimental values are listed. For details see 
    the corresponding references. We found one result by Feng\cite{f04} for
    CoO, obtained by using the B3LYP hybrid 
    functionals method, $J_1= -47.12$ meV and $J_2= -42.56$ meV. To our
    knowledge no further theoretical papers giving numerical values for the
    exchange parameters of FeO exist.} 
  \begin{tabular}{l*{4}{c}l*{3}{c}}
    \hline
    \multicolumn{4}{c}{MnO} & & \multicolumn{4}{c}{NiO}\\
         \cline{1-4}                     \cline{6-9} 
    method &~& $J_1$~~[meV] &~~$J_2$~~[meV]&~~~~&
    method &~& $J_1$~~[meV] &~~$J_2$~~[meV] \\ \hline
    exp.    &        & -2.06,\,-2.64& -2.79& &
    exp.    &        & -0.69,\,0.69 & -8.66,\,-9.51  \\
    this work &      & -0.91        & -1.99& &         %MnO
    this work &      &  0.15        & -6.92          \\%NiO
    LDA+U\cite{st98}&& -2.50        & -6.60& &         %MnO    
    GGA+U\cite{zhh06}&& 0.87        & -9.54          \\%NiO
    OEP\cite{st98}   && -2.85       & -5.50    &&      %MnO
    SIC-LMTO\cite{Diemo_NiO-Paper}&&0.90& -5.50      \\%NiO
    PBE+U\cite{franchini:045132}&&-2.21 & -1.16&&      %MnO
    Fock35\cite{PhysRevB.65.155102}&&0.95&-9.35      \\%NiO
    PBE0\cite{franchini:045132}&& -3.10 & -3.69&&      %MnO
    B3LYP\cite{PhysRevB.65.155102}&&1.20& -13.35     \\%NiO
    HF\cite{franchini:045132}  && -0.73 & -1.16&&      %MnO
    UHF\cite{PhysRevB.65.155102}&& 0.40 & -2.30      \\%NiO
    B3LYP\cite{f04}            && -2.64 & -5.52 &&&& \\\hline
  \end{tabular}

\end{table*}
From our results in Table \ref{tab:J_series} one finds that the $J_2$
parameters constitute 
the major part of magnetic exchange in TMOs and that in magnitude they agree
reasonably well between the two theoretical approaches, MFT and $\Delta
E$. The results
are about 70 to 80 \% of the experimental values, except for FeO, where the
agreement for the MFT $J_2$ is almost perfect and the $\Delta E$ value is
larger than the experimental one. This rather accidental agreement for FeO can
most likely be attributed to the fact that the experimental values are
measured for W\"ustite samples Fe$_{1-x}$O with $x \not=
0$.\cite{khc78}
Also, the experimentally observed trend of the increasing absolute value of
$J_2$ across the series is present in both approaches and is most likely
associated with the increasing number of the TM 3$d$ electrons, responsible
for the magnetic super-exchange.  
Regarding the quantitative agreement with experiment, the MFT results agree
better on average. 
One could envisage that this agreement could be further improved, 
if the MFT approach was applied in the DLM
state\cite{PhysRevB.72.104437,DLM-paper}.  
Nevertheless, compared to the other calculations displayed
in Tables \ref{tab:J_series} and \ref{tab:J_single}, our present results 
may already be considered as being at least as good as those.\\
The situation is very different for the $J_1$ exchange parameters. From Table
\ref{tab:J_series} we see that, with the exception of NiO, the absolute
magnitudes of $J_1$ are, within about 30\%, similar between the two
theoretical approaches, but the signs are opposite. Looking more closely at
the parameters calculated with the MFT we see that the results
show the opposite trend  to that found for $J_{2}$. Namely, the
antiferromagnetic coupling is getting weaker as one moves from MnO to CoO, and
in NiO it becomes ferromagnetic. This can be explained by assuming both kinds
of interaction to be present and to be competing, in the direct and indirect
exchange between NN TM atoms. The picture is relatively intuitive for the
direct case. For Mn, which has half filled $d$-shells, one expects
antiferromagnetic coupling since an electron hopping from one Mn atom to the
other one keeps its spin. Thus, this transfer clearly prefers
antiferromagnetic alignment of the Mn
atoms\cite{Goodenough_Magnetism_book}. Moving across the TMO series the 
occupation of the minority spin channels is growing. This increases the
probability of an electron hopping, if the TM atoms are ferromagnetically
aligned. Thus, the character of the exchange should go towards ferromagnetic,
which is what we find for the $J_1$ calculated via the MFT. Regarding the
indirect exchange, we can also follow 
Goodenough\cite{Goodenough_Magnetism_book}. Nearest TM neighbours interact
antiferromagnetically when two electrons in the 
same oxygen $p$- or $s$-orbitals are excited to the empty TM $e_g$-orbitals. The
strength of this kind of interaction can be assumed not to change a lot along
the TMO series, since the occupation of both the oxygen $p$- or $s$- and the TM
$e_g$-orbitals does not change either. Ferromagnetic coupling on the other
hand is provided by 
electrons of alike spin that are in different orbitals of the O atom. It is
strengthened by a growing occupation of $t_{2g}$-orbitals because this
increases intraatomic exchange. Since the $t_{2g}$-occupancy is
rising when moving across the TMO series from MnO to NiO
\cite{TMO-SIC-paper} one would expect that the magnitude of the
ferromagnetic interaction increases while the antiferromagnetic does not. This
tendency is clearly present in the MFT values for $J_1$ in Table
\ref{tab:J_series}.\\
Looking at the agreement with experiment not much overlap can be spotted. For
the MnO the agreement is satisfactory considering the simplicity of our
Hamiltonian. For FeO the sign is opposite. This could be caused by the
above mentioned fact that in experiment W\"ustite samples of
the kind Fe$_{1-x}$O are investigated while our calculations are performed
for the ideal FeO system. For CoO and NiO comparison is difficult
since experimental values of opposite signs, but similar absolute magnitude
have been measured in different experiments. To conclude the comparison of
MFT-$J_1$ and experimental $J_1$ 
one can say that the agreement is not as good as for the $J_2$. Possible
reasons have just been given, but it also seems that the experimental
determinination of the $J_1$ is not as accurate as for the $J_2$, as can be
seen from the variety of numbers obtained for the same compounds. The lack of
agreement between  
experimental and MFT values for the $J_1$ may influence the calculations based
on those. For the magnon spectra it could be expected that, besides quantitative
differences, due to the different signs even qualitative changes of the
curves might occur. We shall see later, however, that the latter is not the
case. The effect on the calculated N\'eel temperatures should be small in any
case since the energy contributions of the NN in the AFII structure are
canceled out.\\ 
For the $J_1$ calculated with the energy difference approach no obvious 
trends are seen in Table \ref{tab:J_series}, and in addition they are positive 
for all TMOs. For NiO the latter agrees qualitatively with the MFT-$J_1$ and
also with previous theoretical results, and the agreement with those by
K\"odderitzsch \textit{et al.}\cite{Diemo_NiO-Paper}, $J_1=0.9$~meV and $J_2 =
-5.5$~meV, is also quantitatively rather good. For the other TMOs
the sign of $J_1$ is opposite to the ones calculated with the MFT, and 
for MnO and CoO they also do not agree with previous theoretical
investigations. 
The totally different behaviour --- compared to the MFT-values --- can be
explained by looking at the electronic ground states 
of the calculated AFI and FM structures. In both of them the oxygen atoms
carry a magnetic moment, which they do not in the AFII structure. This
magnetic moment can be assumed to give rise to 
magnetic interaction with the neighbouring TM atoms (as a matter of fact,
applying the MFT to AFI or FM structures yields NN exchange parameters of
several meV in magnitude between TM and oxygen atoms). This, however, is in
strong contradiction to the assumptions underlying Eqs.~(\ref{eq:J1}) and
(\ref{eq:J2}), stating that magnetic interaction only occurs
between TM atoms. Thus, when using these equations anyway, this
``artificially'' created 
magnetic exchange is projected onto the $J_1$ and $J_2$. The reason for the
latter quantity being relatively close to its counterpart calculated with the
MFT is probably 
due to the large energy differences between the AFII configuration 
and the AFI as well as the FM configuration for each of the TMOs. This
obviously reduces the error made in Eq.~(\ref{eq:J2}).\\
It should be mentioned that for all
calculated pairs of $J_1$ and $J_2$, using SIC-LSDA, the resulting ground state
magnetic structure is that of AFII\cite{sg88}, despite the relatively large
spread of the $J_1$ parameters. Note that the exchange
  parameters 
$J_1$ and $J_2$, obtained for the TMO series by applying the MFT approach to 
the LSDA ground state electronic structure,
show no agreement with experiment, except for NiO, which can perhaps be
considered as a lucky coincidence. Furthermore, the %$J_{ij}$ 
$J$'s are longer-ranged,
i.e.\ their character is more metallic. This agrees with the fact that their
uncorrected (no SIC) ground states show only very small or no band gaps at all
\cite{TMO-SIC-paper}.
\\
Finally, we would like to comment on the variation of the exchange parameters
as a function of lattice constants shown in Fig.~\ref{fig:J_all} for all the
TMOs.
As one can see, the absolute value of $J_2$ increases
with decreasing $a$. This is in good agreement with the interpretation of the
exchange parameters in terms of overlap integrals. The closer the atoms are,
the larger 
the overlap is between the TM $d$-orbitals and the oxygen $p$-orbitals.
A similar behaviour is found for the $J_1$, which can be understood in the same
way as for $J_2$. However, going through the TMO series and starting with MnO,
the change of the $J_1$ gets smaller as the antiferromagnetic character
becomes less pronounced. According to Goodenough's arguments
\cite{Goodenough_Magnetism_book}, this suggests that the ferromagnetic
coupling becomes more prominent than the antiferromagnetic one. 

%-------------------------------------------------------------------

\subsection{N\'eel temperatures}

\begin{figure}[htbp]
  \centering
  \includegraphics[width=8cm]{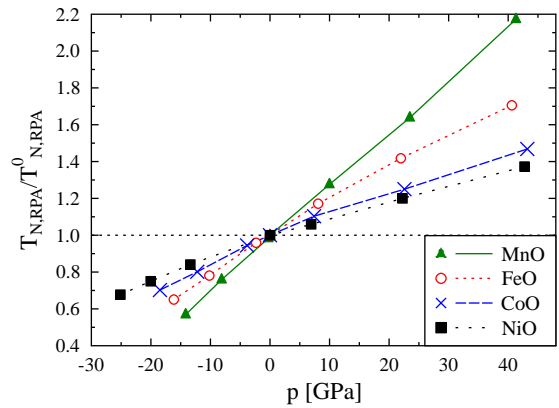}
  \caption{\label{fig:T_N-all} The normalized RPA-based N\'eel temperatures
    $T_{N,\mathrm{RPA}}/T_{N,\mathrm{RPA}}^0$ for all TMOs as a function of pressure
    $p$. $T_{N,\mathrm{RPA}}^0$ is taken from row 3 of Table
    \ref{tab:T_N-summary}.} 
\end{figure}

The calculated transition temperatures are summarized in Table
\ref{tab:T_N-summary}. 
\begin{table}[htbp]
  \caption{\label{tab:T_N-summary}Summary of the N\'eel temperatures 
    calculated with the $J_{ij}$ from the MFT approach (see Table
    \ref{tab:J_series}). 
    In the top two rows the experimental and the DLM values are listed,
    followed by the RPA values based on the interaction of the first 11
    TM-TM-shells and of only the nearest and next-nearest neighbours
    (i.e. only $J_1$ and $J_2$). In rows 5 and 6 the MFA results shown, again
    using 11 or 2 shells, respectively. In the last row the results of the
    Monte Carlo simulations are presented.}
  \begin{tabular}{l*{5}{r}} \hline
    \multicolumn{1}{c}{$T_N$  [K]} &&
    \multicolumn{1}{c}{~MnO} &
    \multicolumn{1}{c}{~FeO} &
    \multicolumn{1}{c}{~CoO} &
    \multicolumn{1}{c}{~NiO}  \\ \hline \hline
    Experiment        && 118 & 192 & 289 & 523  \\\hline %\hline
    DLM\cite{DLM-paper}&& 126 & 172 & 242 & 336 \\\hline
    RPA with $J_{1-11}$ &&  81 & 146 & 252 & 440  \\%\hline %\hline
    RPA with $J_{1,2}$ &&  87 & 155 & 260 & 448  \\ \hline
    MFA with $J_{1-11}$ && 122 & 210 & 362 & 628  \\%\hline
    MFA with $J_{1,2}$ && 129 & 221 & 373 & 644  \\ \hline
    MC                &&  90 & 162 & 260 & 458  \\%\hline %\hline
  \hline
  \end{tabular}
\end{table}
One finds that MFA overestimates the experimental N\'eel
temperatures, whereas RPA underestimates them. This is what can be expected
from general considerations\cite{Tkrit-RPA-Rusz}. 
One can also see that the N\'eel temperatures calculated in the RPA approach
based only on $J_1$ and $J_2$ do not differ significantly from those
calculated using 
the 11 neighbour shells. This again agrees with the idea of superexchange. 
What is not expected is that, for MnO and FeO the RPA and MC results are 
relatively small compared to experiment. In fact, for these TMOs the MFA gives
a better estimate. The 
probable reason for that is the general relative underestimate for the $J_2$.
The latter being the main contribution of magnetic exchange, their underestimate
is largest for MnO and decreases towards NiO. An exception is FeO. The
agreement for $J_2$ is almost perfect, yet the RPA and MC estimates are roughly
of the same quality as those for the other TMOs. However, it can again
(see discussion of $J_1$ and $J_2$) be argued that due to 
the experimental imperfect FeO lattice other effects not considered in our
approach may play an important role for the formation of magnetic order. 
The DLM results of Hughes \textit{et al.}\cite{DLM-paper} are, with the
exception of NiO, in good agreement with the experimental values. Their
trend, however, is opposite to ours, namely the ratio
$T_N^{\mathrm{DLM}}/T_N^{\mathrm{exp}}$ becomes smaller with increasing atomic
number. This could be due to not taking into account the quantum character of
the systems, which in the present paper is done via the factor
$(S+1)/S$\cite{QuantumScaling_note}, where $S$ is calculated according to
Hund's rules. Another possible reason especially for
the NiO result, as discussed in
Ref.~\onlinecite{DLM-paper}, might be related to a possible importance of the
short range order correlations that a single-site approximation like DLM would
not do justice to.\\
Concentrating on our RPA and MC results, we have to admit that better
calculations for the individually selected TMO systems can be found in
literature. Among them are the calculations by Zhang \textit{et
  al.}\cite{zhh06} for NiO (with a rather semi-empirical approach) and Towler
\textit{et al.}\cite{tah94} 
for MnO. However, when studying the whole TMO series with the same approach,
such as the above DLM application, the work reported by Harrison \cite{h07} or
Wan, Yin and Savrasov \cite{wys06}, it is hard to find \textit{ab initio}
results, treating electron correlations at the same level of  
sophistication and predicting the N\'eel temperatures qualitatively and
quantitatively as accurately as in the present paper throughout the whole TMO
series.\\
To finish, we briefly discuss the N\'eel temperature dependence on pressure
shown in Fig.~\ref{fig:T_N-all} for all the studied TMOs.
To calculate the pressure, $p$, the Murnaghan equation of state\cite{Murnaghan}
has been used. 
Based on the behaviour of $J_2$ seen in Fig.~\ref{fig:J_all}, it
is not surprising that for the whole TMO series the calculated N\'eel
temperatures increase with pressure. 
Qualitatively, this agrees with previous experimental and theoretical results,
indicating a stability of the antiferromagnetic structure up to high
pressures (several tens of GPa, at least) before it collapses and a
paramagnetic or low spin configuration takes over.
\cite{ymk05,ding:144101,PhysRev.184.323,bloch:1401,kasinathan:195110,sidorov:2174,JPSJ.23.1174,PhysRevB.47.7720,bss99,zhh06}.
We can compare the pressure dependency of $T_N$ to
experiment (for MnO -- Ref.~\onlinecite{ymk05}, FeO --
Ref.~\onlinecite{JPSJ.23.1174}, CoO -- Ref.~\onlinecite{PhysRev.184.323}, NiO --
Ref.~\onlinecite{sidorov:2174}) for $p>0$ by assuming them to be
linear. Taking the pressure  
derivative of the normalized N\'eel temperatures, $\partial (T_N / T_N(p=0)) /
\partial p$, we find that our calculated values increase too slowly, roughly
by a factor of 1/2. 

%-------------------------------------------------------------------

\subsection{Magnon Spectra}

\begin{figure}[htbp]
  \centering
  \includegraphics[width=8.5cm]{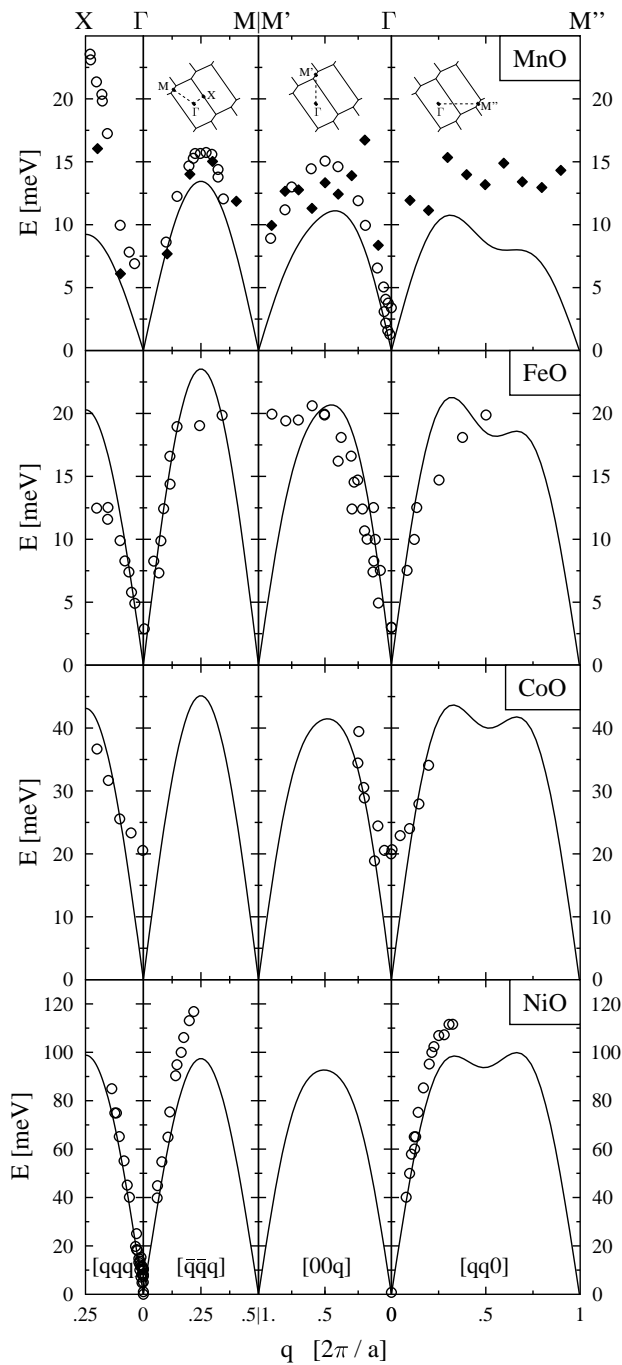}
    \caption{\label{fig:Magnons_all}
      Shown are the calculated TMO spin wave dispersions together with 
      experimental data points for MnO (black diamonds\cite{gdt07}, open
      circles\cite{p74}), FeO\cite{khc78}, CoO\cite{ti06} 
      and NiO\cite{Hutchings_Samuelson}, respectively. The coordinates are
      cartesian and in units of $2\pi / a$. The path chosen along several high
      symmetry lines starts at X$=(0.25,0.25,0.25)$ and goes along $[qqq]$ 
      to $\Gamma = (0,0,0)$, then
      along $[\overline{q}\overline{q}q]$ to M$=(-0.5,-0.5,0.5)$, and further 
      along $[00q]$ to $\Gamma$ of the neighbouring AFII Brillouin zone, then
      continuing along $[qq0]$ to M''. The inlays in the MnO panel show the
      different branches along the AFII Brillouin zone. 
    }
\end{figure}
Considering the above results for the $J_{ij}$ parameters it is reasonable to
assume 
that only the nearest and next-nearest neighbour interactions contribute
significantly to the magnon dispersion relation, which therefore should be
adequately represented by Eq.~(\ref{eq:Terakuraformel}). For the calculation
the MFT-$J_1$ and $J_2$ from Table \ref{tab:J_series} and the theoretical (calc.)
magnetic moment $\mu$ from Table \ref{tab:a0} were used. The resulting magnon
spectra for all the studied TMOs, in the AFII structure, are shown in
Fig.~\ref{fig:Magnons_all} together with the experimental results. 
Generally, the agreement between the calculated dispersion curves and the
experimental observations is rather good, considering the 
Heisenberg Hamiltonian used in this work --- anisotropy and alignment energy
terms are neglected. This is also the reason why the calculated curves fail to
reproduce 
the non-zero energies at M$=(-0.5,-0.5,0.5)$. Besides that, minima, maxima and
curvature are well reproduced. Furthermore it can be seen that except for
FeO the theoretical curves generally underestimate the experimental energies,
which is due to the underestimate of the $J_2$ 
parameters. The relative magnitude of the peak along $[qqq]$ varies strongly,
as one goes through the TMO series. This effect can be ascribed to the
changing ratio of $J_2/J_1$.\\
The qualitative agreement with previous theoretical works, e.\ g.\ such as that
of Solovyev and Terakura\cite{st98} is good, although not in the absolute
numerical terms, arising from different values of the Heisenberg exchange
parameters $J_{ij}$.

\section{Conclusion \label{sec:conclusion}}
We have used the local self-interaction correction, implemented in
the  multiple scattering theory in the framework of KKR in combination with
the magnetic  force theorem to study magnetic interactions in transition metal
monoxides. Specifically, we have calculated the $J_{1}$ and $J_{2}$ exchange
parameters, the corresponding N\'eel temperatures and the respective magnon
spectra for the whole TMO series. The most important conclusion of this work
is that the combined approach used here provides an adequate
description of magnetic interactions for the series as a whole. Without
considering correlation effects the theoretical results in general do not
agree with 
experimental findings. Furthermore, we have shown that our ab-initio approach
yields upper (MFA) and lower limits (RPA, MC simulations) for the
N\'eel temperatures for the whole TMO series, and the calculated magnon  
spectra are in good qualitative agreement with experiment and other
theoretical calculations.

\section*{Acknowledgements}
We would like to thank Julie Staunton for helpful discussions and comments.
This work was supported by the Deutsche Forschungsgesellschaft (DFG) via the
SFB 762 ``Functionality of Oxidic Interfaces''. Calculations were performed at
the John von Neumann Institute for Computing in J\"ulich, Germany. 
Research at the Oak Ridge National Laboratory was sponsored by the Division of
Materials Sciences and Engineering, Office of Basic Energy Sciences, US
Department of Energy, under Contract DE-AC05-00OR22725 with UT-Battelle, LLC.

\bibliography{tmopaper}

\begin{thebibliography}{69}
\expandafter\ifx\csname natexlab\endcsname\relax\def\natexlab#1{#1}\fi
\expandafter\ifx\csname bibnamefont\endcsname\relax
  \def\bibnamefont#1{#1}\fi
\expandafter\ifx\csname bibfnamefont\endcsname\relax
  \def\bibfnamefont#1{#1}\fi
\expandafter\ifx\csname citenamefont\endcsname\relax
  \def\citenamefont#1{#1}\fi
\expandafter\ifx\csname url\endcsname\relax
  \def\url#1{\texttt{#1}}\fi
\expandafter\ifx\csname urlprefix\endcsname\relax\def\urlprefix{URL }\fi
\providecommand{\bibinfo}[2]{#2}
\providecommand{\eprint}[2][]{\url{#2}}

\bibitem[{\citenamefont{Struzhkin et~al.}(2001)\citenamefont{Struzhkin, Mao,
  Hu, Schwoerer-B\"ohning, Shu, Hemley, Sturhahn, Hu, Alp, Eng
  et~al.}}]{PhysRevLett.87.255501}
\bibinfo{author}{\bibfnamefont{V.~V.} \bibnamefont{Struzhkin}},
  \bibinfo{author}{\bibfnamefont{H.-k.} \bibnamefont{Mao}},
  \bibinfo{author}{\bibfnamefont{J.}~\bibnamefont{Hu}},
  \bibinfo{author}{\bibfnamefont{M.}~\bibnamefont{Schwoerer-B\"ohning}},
  \bibinfo{author}{\bibfnamefont{J.}~\bibnamefont{Shu}},
  \bibinfo{author}{\bibfnamefont{R.~J.} \bibnamefont{Hemley}},
  \bibinfo{author}{\bibfnamefont{W.}~\bibnamefont{Sturhahn}},
  \bibinfo{author}{\bibfnamefont{M.~Y.} \bibnamefont{Hu}},
  \bibinfo{author}{\bibfnamefont{E.~E.} \bibnamefont{Alp}},
  \bibinfo{author}{\bibfnamefont{P.}~\bibnamefont{Eng}}, \bibnamefont{et~al.},
  \bibinfo{journal}{Phys. Rev. Lett.} \textbf{\bibinfo{volume}{87}},
  \bibinfo{pages}{255501} (\bibinfo{year}{2001}).

\bibitem[{\citenamefont{Lines and Jones}(1965)}]{l652}
\bibinfo{author}{\bibfnamefont{M.}~\bibnamefont{Lines}} \bibnamefont{and}
  \bibinfo{author}{\bibfnamefont{E.}~\bibnamefont{Jones}},
  \bibinfo{journal}{Phys. Rev.} \textbf{\bibinfo{volume}{139}},
  \bibinfo{pages}{A1313} (\bibinfo{year}{1965}).

\bibitem[{\citenamefont{Patterson et~al.}(2004)\citenamefont{Patterson, Aracne,
  Jackson, Malba, Weir, Baker, and Vohra}}]{p04}
\bibinfo{author}{\bibfnamefont{J.~R.} \bibnamefont{Patterson}},
  \bibinfo{author}{\bibfnamefont{C.~M.} \bibnamefont{Aracne}},
  \bibinfo{author}{\bibfnamefont{D.~D.} \bibnamefont{Jackson}},
  \bibinfo{author}{\bibfnamefont{V.}~\bibnamefont{Malba}},
  \bibinfo{author}{\bibfnamefont{S.~T.} \bibnamefont{Weir}},
  \bibinfo{author}{\bibfnamefont{P.~A.} \bibnamefont{Baker}}, \bibnamefont{and}
  \bibinfo{author}{\bibfnamefont{Y.~K.} \bibnamefont{Vohra}},
  \bibinfo{journal}{Phys. Rev. B} \textbf{\bibinfo{volume}{69}},
  \bibinfo{pages}{220101(R)} (\bibinfo{year}{2004}).

\bibitem[{\citenamefont{Yoo et~al.}(2005)\citenamefont{Yoo, Maddox, Klepeis,
  Iota, Evans, McMahan, Hu, Chow, Somayazulu, Hausermann et~al.}}]{ymk05}
\bibinfo{author}{\bibfnamefont{C.~S.} \bibnamefont{Yoo}},
  \bibinfo{author}{\bibfnamefont{B.}~\bibnamefont{Maddox}},
  \bibinfo{author}{\bibfnamefont{J.~H.~P.} \bibnamefont{Klepeis}},
  \bibinfo{author}{\bibfnamefont{V.}~\bibnamefont{Iota}},
  \bibinfo{author}{\bibfnamefont{W.}~\bibnamefont{Evans}},
  \bibinfo{author}{\bibfnamefont{A.}~\bibnamefont{McMahan}},
  \bibinfo{author}{\bibfnamefont{M.~Y.} \bibnamefont{Hu}},
  \bibinfo{author}{\bibfnamefont{P.}~\bibnamefont{Chow}},
  \bibinfo{author}{\bibfnamefont{M.}~\bibnamefont{Somayazulu}},
  \bibinfo{author}{\bibfnamefont{D.}~\bibnamefont{Hausermann}},
  \bibnamefont{et~al.}, \bibinfo{journal}{Physical Review Letters}
  \textbf{\bibinfo{volume}{94}}, \bibinfo{pages}{115502}
  (\bibinfo{year}{2005}).

\bibitem[{\citenamefont{Mattila et~al.}(2007)\citenamefont{Mattila, Rueff,
  Badro, Vanko, and Shukla}}]{mrb07}
\bibinfo{author}{\bibfnamefont{A.}~\bibnamefont{Mattila}},
  \bibinfo{author}{\bibfnamefont{J.-P.} \bibnamefont{Rueff}},
  \bibinfo{author}{\bibfnamefont{J.}~\bibnamefont{Badro}},
  \bibinfo{author}{\bibfnamefont{G.}~\bibnamefont{Vanko}}, \bibnamefont{and}
  \bibinfo{author}{\bibfnamefont{A.}~\bibnamefont{Shukla}},
  \bibinfo{journal}{Physical Review Letters} \textbf{\bibinfo{volume}{98}},
  \bibinfo{pages}{196404} (\bibinfo{year}{2007}).

\bibitem[{\citenamefont{Kasinathan et~al.}(2006)\citenamefont{Kasinathan,
  Kunes, Koepernik, Diaconu, Martin, Prodan, Scuseria, Spaldin, Petit,
  Schulthess et~al.}}]{kasinathan:195110}
\bibinfo{author}{\bibfnamefont{D.}~\bibnamefont{Kasinathan}},
  \bibinfo{author}{\bibfnamefont{J.}~\bibnamefont{Kunes}},
  \bibinfo{author}{\bibfnamefont{K.}~\bibnamefont{Koepernik}},
  \bibinfo{author}{\bibfnamefont{C.~V.} \bibnamefont{Diaconu}},
  \bibinfo{author}{\bibfnamefont{R.~L.} \bibnamefont{Martin}},
  \bibinfo{author}{\bibfnamefont{I.~D.} \bibnamefont{Prodan}},
  \bibinfo{author}{\bibfnamefont{G.~E.} \bibnamefont{Scuseria}},
  \bibinfo{author}{\bibfnamefont{N.}~\bibnamefont{Spaldin}},
  \bibinfo{author}{\bibfnamefont{L.}~\bibnamefont{Petit}},
  \bibinfo{author}{\bibfnamefont{T.~C.} \bibnamefont{Schulthess}},
  \bibnamefont{et~al.}, \bibinfo{journal}{Phys. Rev. B}
  \textbf{\bibinfo{volume}{74}}, \bibinfo{eid}{195110}
  (pages~\bibinfo{numpages}{12}) (\bibinfo{year}{2006}).

\bibitem[{\citenamefont{Wdowik and Legut}(2008)}]{wl08}
\bibinfo{author}{\bibfnamefont{U.}~\bibnamefont{Wdowik}} \bibnamefont{and}
  \bibinfo{author}{\bibfnamefont{D.}~\bibnamefont{Legut}}, \bibinfo{journal}{J.
  Phys. Chem. Sol.} \textbf{\bibinfo{volume}{69}}, \bibinfo{pages}{1698}
  (\bibinfo{year}{2008}).

\bibitem[{\citenamefont{Goodwin et~al.}(2005)\citenamefont{Goodwin, Tucker,
  Cope, Dove, and Keen}}]{gtc05}
\bibinfo{author}{\bibfnamefont{A.~L.} \bibnamefont{Goodwin}},
  \bibinfo{author}{\bibfnamefont{M.~G.} \bibnamefont{Tucker}},
  \bibinfo{author}{\bibfnamefont{E.~R.} \bibnamefont{Cope}},
  \bibinfo{author}{\bibfnamefont{M.~T.} \bibnamefont{Dove}}, \bibnamefont{and}
  \bibinfo{author}{\bibfnamefont{D.~A.} \bibnamefont{Keen}},
  \bibinfo{journal}{Phys. Rev. B} \textbf{\bibinfo{volume}{72}},
  \bibinfo{pages}{214304} (\bibinfo{year}{2005}).

\bibitem[{\citenamefont{Goodwin et~al.}(2006)\citenamefont{Goodwin, Tucker,
  Dove, and Keen}}]{gtd06}
\bibinfo{author}{\bibfnamefont{A.~L.} \bibnamefont{Goodwin}},
  \bibinfo{author}{\bibfnamefont{M.~G.} \bibnamefont{Tucker}},
  \bibinfo{author}{\bibfnamefont{M.~T.} \bibnamefont{Dove}}, \bibnamefont{and}
  \bibinfo{author}{\bibfnamefont{D.~A.} \bibnamefont{Keen}},
  \bibinfo{journal}{Physical Review Letters} \textbf{\bibinfo{volume}{96}},
  \bibinfo{eid}{047209} (pages~\bibinfo{numpages}{4}) (\bibinfo{year}{2006}).

\bibitem[{\citenamefont{Goodwin et~al.}(2007)\citenamefont{Goodwin, Dove,
  Tucker, and Keen}}]{gdt07}
\bibinfo{author}{\bibfnamefont{A.~L.} \bibnamefont{Goodwin}},
  \bibinfo{author}{\bibfnamefont{M.~T.} \bibnamefont{Dove}},
  \bibinfo{author}{\bibfnamefont{M.~G.} \bibnamefont{Tucker}},
  \bibnamefont{and} \bibinfo{author}{\bibfnamefont{D.~A.} \bibnamefont{Keen}},
  \bibinfo{journal}{Phys. Rev. B} \textbf{\bibinfo{volume}{75}},
  \bibinfo{eid}{075423} (pages~\bibinfo{numpages}{9}) (\bibinfo{year}{2007}).

\bibitem[{\citenamefont{Ott}(2008)}]{o08}
\bibinfo{author}{\bibfnamefont{F.}~\bibnamefont{Ott}}, \bibinfo{journal}{J.
  Phys.: Condens. Matter} \textbf{\bibinfo{volume}{20}},
  \bibinfo{pages}{264009} (\bibinfo{year}{2008}).

\bibitem[{\citenamefont{van~der Zaag et~al.}(2000)\citenamefont{van~der Zaag,
  Ijiri, Borchers, Feiner, Wolf, Gaines, Erwin, and Verheijen}}]{zib00}
\bibinfo{author}{\bibfnamefont{P.~J.} \bibnamefont{van~der Zaag}},
  \bibinfo{author}{\bibfnamefont{Y.}~\bibnamefont{Ijiri}},
  \bibinfo{author}{\bibfnamefont{J.~A.} \bibnamefont{Borchers}},
  \bibinfo{author}{\bibfnamefont{L.~F.} \bibnamefont{Feiner}},
  \bibinfo{author}{\bibfnamefont{R.~M.} \bibnamefont{Wolf}},
  \bibinfo{author}{\bibfnamefont{J.~M.} \bibnamefont{Gaines}},
  \bibinfo{author}{\bibfnamefont{R.~W.} \bibnamefont{Erwin}}, \bibnamefont{and}
  \bibinfo{author}{\bibfnamefont{M.~A.} \bibnamefont{Verheijen}},
  \bibinfo{journal}{Phys. Rev. Lett.} \textbf{\bibinfo{volume}{84}},
  \bibinfo{pages}{6102} (\bibinfo{year}{2000}).

\bibitem[{\citenamefont{Bengone et~al.}(2000)\citenamefont{Bengone, Alouani,
  Bl\"ochl, and Hugel}}]{PhysRevB.62.16392}
\bibinfo{author}{\bibfnamefont{O.}~\bibnamefont{Bengone}},
  \bibinfo{author}{\bibfnamefont{M.}~\bibnamefont{Alouani}},
  \bibinfo{author}{\bibfnamefont{P.}~\bibnamefont{Bl\"ochl}}, \bibnamefont{and}
  \bibinfo{author}{\bibfnamefont{J.}~\bibnamefont{Hugel}},
  \bibinfo{journal}{Phys. Rev. B} \textbf{\bibinfo{volume}{62}},
  \bibinfo{pages}{16392} (\bibinfo{year}{2000}).

\bibitem[{\citenamefont{Rohrbach et~al.}(2004)\citenamefont{Rohrbach, Hafner,
  and Kresse}}]{PhysRevB.69.075413}
\bibinfo{author}{\bibfnamefont{A.}~\bibnamefont{Rohrbach}},
  \bibinfo{author}{\bibfnamefont{J.}~\bibnamefont{Hafner}}, \bibnamefont{and}
  \bibinfo{author}{\bibfnamefont{G.}~\bibnamefont{Kresse}},
  \bibinfo{journal}{Phys. Rev. B} \textbf{\bibinfo{volume}{69}},
  \bibinfo{pages}{075413} (\bibinfo{year}{2004}).

\bibitem[{\citenamefont{Zhang et~al.}(2006)\citenamefont{Zhang, Hu, Han, and
  Tang}}]{zhh06}
\bibinfo{author}{\bibfnamefont{W.-B.} \bibnamefont{Zhang}},
  \bibinfo{author}{\bibfnamefont{Y.-L.} \bibnamefont{Hu}},
  \bibinfo{author}{\bibfnamefont{K.-L.} \bibnamefont{Han}}, \bibnamefont{and}
  \bibinfo{author}{\bibfnamefont{B.-Y.} \bibnamefont{Tang}},
  \bibinfo{journal}{Phys. Rev. B} \textbf{\bibinfo{volume}{74}},
  \bibinfo{pages}{054421} (\bibinfo{year}{2006}).

\bibitem[{\citenamefont{Temmerman et~al.}(1998)\citenamefont{Temmerman, Svane,
  Szotek, and Winter}}]{Temmerman98}
\bibinfo{author}{\bibfnamefont{W.~M.} \bibnamefont{Temmerman}},
  \bibinfo{author}{\bibfnamefont{A.}~\bibnamefont{Svane}},
  \bibinfo{author}{\bibfnamefont{Z.}~\bibnamefont{Szotek}}, \bibnamefont{and}
  \bibinfo{author}{\bibfnamefont{H.}~\bibnamefont{Winter}}, in
  \emph{\bibinfo{booktitle}{Electronic Density Functional Theory: Recent
  Progress and New Directions}}, edited by
  \bibinfo{editor}{\bibfnamefont{J.~F.} \bibnamefont{Dobson}},
  \bibinfo{editor}{\bibfnamefont{G.}~\bibnamefont{Vignale}}, \bibnamefont{and}
  \bibinfo{editor}{\bibfnamefont{M.~P.} \bibnamefont{Das}}
  (\bibinfo{publisher}{Plenum}, \bibinfo{address}{New York},
  \bibinfo{year}{1998}), p. \bibinfo{pages}{327}.

\bibitem[{\citenamefont{Temmerman et~al.}(2000)\citenamefont{Temmerman, Svane,
  Szotek, Winter, and Beiden}}]{tss00}
\bibinfo{author}{\bibfnamefont{W.}~\bibnamefont{Temmerman}},
  \bibinfo{author}{\bibfnamefont{A.}~\bibnamefont{Svane}},
  \bibinfo{author}{\bibfnamefont{Z.}~\bibnamefont{Szotek}},
  \bibinfo{author}{\bibfnamefont{H.}~\bibnamefont{Winter}}, \bibnamefont{and}
  \bibinfo{author}{\bibfnamefont{S.}~\bibnamefont{Beiden}}, in
  \emph{\bibinfo{booktitle}{Electronic Structure and Physical Properties of
  Solids - The use of the LMTO Method}} (\bibinfo{publisher}{Springer},
  \bibinfo{address}{Berlin Heidelberg New York}, \bibinfo{year}{2000}), Lecture
  notes in Physics.

\bibitem[{\citenamefont{Svane and Gunnarsson}(1990)}]{sg90}
\bibinfo{author}{\bibfnamefont{A.}~\bibnamefont{Svane}} \bibnamefont{and}
  \bibinfo{author}{\bibfnamefont{O.}~\bibnamefont{Gunnarsson}},
  \bibinfo{journal}{Physical Review Letters} \textbf{\bibinfo{volume}{65}},
  \bibinfo{pages}{1148} (\bibinfo{year}{1990}).

\bibitem[{\citenamefont{Szotek et~al.}(1993)\citenamefont{Szotek, Temmerman,
  and Winter}}]{stw93}
\bibinfo{author}{\bibfnamefont{Z.}~\bibnamefont{Szotek}},
  \bibinfo{author}{\bibfnamefont{W.~M.} \bibnamefont{Temmerman}},
  \bibnamefont{and} \bibinfo{author}{\bibfnamefont{H.}~\bibnamefont{Winter}},
  \bibinfo{journal}{Phys. Rev. B} \textbf{\bibinfo{volume}{47}},
  \bibinfo{pages}{4029} (\bibinfo{year}{1993}).

\bibitem[{\citenamefont{K\"odderitzsch
  et~al.}(2002)\citenamefont{K\"odderitzsch, Hergert, Temmerman, Szotek, Ernst,
  and Winter}}]{Diemo_NiO-Paper}
\bibinfo{author}{\bibfnamefont{D.}~\bibnamefont{K\"odderitzsch}},
  \bibinfo{author}{\bibfnamefont{W.}~\bibnamefont{Hergert}},
  \bibinfo{author}{\bibfnamefont{W.~M.} \bibnamefont{Temmerman}},
  \bibinfo{author}{\bibfnamefont{Z.}~\bibnamefont{Szotek}},
  \bibinfo{author}{\bibfnamefont{A.}~\bibnamefont{Ernst}}, \bibnamefont{and}
  \bibinfo{author}{\bibfnamefont{H.}~\bibnamefont{Winter}},
  \bibinfo{journal}{Phys. Rev. B} \textbf{\bibinfo{volume}{66}},
  \bibinfo{pages}{064434} (\bibinfo{year}{2002}).

\bibitem[{\citenamefont{D\"ane et~al.}(2009)\citenamefont{D\"ane, L\"uders,
  Ernst, K\"odderitzsch, Temmerman, Szotek, and Hergert}}]{TMO-SIC-paper}
\bibinfo{author}{\bibfnamefont{M.}~\bibnamefont{D\"ane}},
  \bibinfo{author}{\bibfnamefont{M.}~\bibnamefont{L\"uders}},
  \bibinfo{author}{\bibfnamefont{A.}~\bibnamefont{Ernst}},
  \bibinfo{author}{\bibfnamefont{D.}~\bibnamefont{K\"odderitzsch}},
  \bibinfo{author}{\bibfnamefont{W.}~\bibnamefont{Temmerman}},
  \bibinfo{author}{\bibfnamefont{Z.}~\bibnamefont{Szotek}}, \bibnamefont{and}
  \bibinfo{author}{\bibfnamefont{W.}~\bibnamefont{Hergert}},
  \bibinfo{journal}{Journal of Physics: Condensed Matter}
  \textbf{\bibinfo{volume}{21}}, \bibinfo{pages}{045604}
  (\bibinfo{year}{2009}).

\bibitem[{\citenamefont{Franchini et~al.}(2005)\citenamefont{Franchini, Bayer,
  Podloucky, Paier, and Kresse}}]{franchini:045132}
\bibinfo{author}{\bibfnamefont{C.}~\bibnamefont{Franchini}},
  \bibinfo{author}{\bibfnamefont{V.}~\bibnamefont{Bayer}},
  \bibinfo{author}{\bibfnamefont{R.}~\bibnamefont{Podloucky}},
  \bibinfo{author}{\bibfnamefont{J.}~\bibnamefont{Paier}}, \bibnamefont{and}
  \bibinfo{author}{\bibfnamefont{G.}~\bibnamefont{Kresse}},
  \bibinfo{journal}{Phys. Rev. B} \textbf{\bibinfo{volume}{72}},
  \bibinfo{eid}{045132} (pages~\bibinfo{numpages}{6}) (\bibinfo{year}{2005}).

\bibitem[{\citenamefont{Feng}(2004)}]{f04}
\bibinfo{author}{\bibfnamefont{X.}~\bibnamefont{Feng}}, \bibinfo{journal}{Phys.
  Rev. B} \textbf{\bibinfo{volume}{69}}, \bibinfo{pages}{155107}
  (\bibinfo{year}{2004}).

\bibitem[{\citenamefont{Kunes et~al.}(2007)\citenamefont{Kunes, Anisimov,
  Skornyakov, Lukoyanov, and Vollhardt}}]{kas07}
\bibinfo{author}{\bibfnamefont{J.}~\bibnamefont{Kunes}},
  \bibinfo{author}{\bibfnamefont{V.~I.} \bibnamefont{Anisimov}},
  \bibinfo{author}{\bibfnamefont{S.~L.} \bibnamefont{Skornyakov}},
  \bibinfo{author}{\bibfnamefont{A.~V.} \bibnamefont{Lukoyanov}},
  \bibnamefont{and}
  \bibinfo{author}{\bibfnamefont{D.}~\bibnamefont{Vollhardt}},
  \bibinfo{journal}{Physical Review Letters} \textbf{\bibinfo{volume}{99}},
  \bibinfo{pages}{156404} (\bibinfo{year}{2007}).

\bibitem[{\citenamefont{Perdew and Zunger}(1981)}]{pz81}
\bibinfo{author}{\bibfnamefont{J.~P.} \bibnamefont{Perdew}} \bibnamefont{and}
  \bibinfo{author}{\bibfnamefont{A.}~\bibnamefont{Zunger}},
  \bibinfo{journal}{Phys. Rev. B} \textbf{\bibinfo{volume}{23}},
  \bibinfo{pages}{5048} (\bibinfo{year}{1981}).

\bibitem[{\citenamefont{Liechtenstein et~al.}(1987)\citenamefont{Liechtenstein,
  Katsnelson, Antropov, and Gubanov}}]{lka87}
\bibinfo{author}{\bibfnamefont{A.}~\bibnamefont{Liechtenstein}},
  \bibinfo{author}{\bibfnamefont{M.}~\bibnamefont{Katsnelson}},
  \bibinfo{author}{\bibfnamefont{V.}~\bibnamefont{Antropov}}, \bibnamefont{and}
  \bibinfo{author}{\bibfnamefont{V.}~\bibnamefont{Gubanov}},
  \bibinfo{journal}{J. of Mag. Mag. Mat.} \textbf{\bibinfo{volume}{67}},
  \bibinfo{pages}{65} (\bibinfo{year}{1987}).

\bibitem[{\citenamefont{L\"uders et~al.}(2005)\citenamefont{L\"uders, Ernst,
  Dane, Szotek, Svane, Kodderitzsch, Hergert, Gyorffy, and Temmerman}}]{LED05}
\bibinfo{author}{\bibfnamefont{M.}~\bibnamefont{L\"uders}},
  \bibinfo{author}{\bibfnamefont{A.}~\bibnamefont{Ernst}},
  \bibinfo{author}{\bibfnamefont{M.}~\bibnamefont{Dane}},
  \bibinfo{author}{\bibfnamefont{Z.}~\bibnamefont{Szotek}},
  \bibinfo{author}{\bibfnamefont{A.}~\bibnamefont{Svane}},
  \bibinfo{author}{\bibfnamefont{D.}~\bibnamefont{Kodderitzsch}},
  \bibinfo{author}{\bibfnamefont{W.}~\bibnamefont{Hergert}},
  \bibinfo{author}{\bibfnamefont{B.~L.} \bibnamefont{Gyorffy}},
  \bibnamefont{and} \bibinfo{author}{\bibfnamefont{W.~M.}
  \bibnamefont{Temmerman}}, \bibinfo{journal}{Phys. Rev. B}
  \textbf{\bibinfo{volume}{71}}, \bibinfo{pages}{205109}
  (\bibinfo{year}{2005}).

\bibitem[{\citenamefont{Hughes et~al.}(2007)\citenamefont{Hughes, D\"ane,
  Ernst, Hergert, L\"uders, Poulter, Staunton, Svane, Szotek, and
  Temmerman}}]{hde07}
\bibinfo{author}{\bibfnamefont{I.~D.} \bibnamefont{Hughes}},
  \bibinfo{author}{\bibfnamefont{M.}~\bibnamefont{D\"ane}},
  \bibinfo{author}{\bibfnamefont{A.}~\bibnamefont{Ernst}},
  \bibinfo{author}{\bibfnamefont{W.}~\bibnamefont{Hergert}},
  \bibinfo{author}{\bibfnamefont{M.}~\bibnamefont{L\"uders}},
  \bibinfo{author}{\bibfnamefont{J.}~\bibnamefont{Poulter}},
  \bibinfo{author}{\bibfnamefont{J.~B.} \bibnamefont{Staunton}},
  \bibinfo{author}{\bibfnamefont{A.}~\bibnamefont{Svane}},
  \bibinfo{author}{\bibfnamefont{Z.}~\bibnamefont{Szotek}}, \bibnamefont{and}
  \bibinfo{author}{\bibfnamefont{W.~M.} \bibnamefont{Temmerman}},
  \bibinfo{journal}{Nature} \textbf{\bibinfo{volume}{446}},
  \bibinfo{pages}{650} (\bibinfo{year}{2007}).

\bibitem[{\citenamefont{Hughes et~al.}(2008)\citenamefont{Hughes, D\"ane,
  Ernst, Hergert, L\"uders, Staunton, Szotek, and Temmerman}}]{DLM-paper}
\bibinfo{author}{\bibfnamefont{I.}~\bibnamefont{Hughes}},
  \bibinfo{author}{\bibfnamefont{M.}~\bibnamefont{D\"ane}},
  \bibinfo{author}{\bibfnamefont{A.}~\bibnamefont{Ernst}},
  \bibinfo{author}{\bibfnamefont{W.}~\bibnamefont{Hergert}},
  \bibinfo{author}{\bibfnamefont{M.}~\bibnamefont{L\"uders}},
  \bibinfo{author}{\bibfnamefont{J.~B.} \bibnamefont{Staunton}},
  \bibinfo{author}{\bibfnamefont{Z.}~\bibnamefont{Szotek}}, \bibnamefont{and}
  \bibinfo{author}{\bibfnamefont{W.}~\bibnamefont{Temmerman}},
  \bibinfo{journal}{New Journal of Physics} \textbf{\bibinfo{volume}{10}},
  \bibinfo{pages}{063010} (\bibinfo{year}{2008}).

\bibitem[{Ham()}]{Hamiltonian_note}
\bibinfo{note}{The Heisenberg Hamiltonian can be defined in several ways. Often
  the sum is multiplied with the factor 1/2, which corresponds to counting each
  $ij$-pair only once. Sometimes the minus sign is omitted. In our case also
  the absolute values of the spin vectors $\fat{S}_i$ and $\fat{S}_j$ are
  included in the $J_{ij}$ and instead the unit vectors $\fat{e}_i$ and
  $\fat{e}_j$ are used. One has to take care of this when comparing exchange
  parameters $J_{ij}$ of different works.}

\bibitem[{\citenamefont{Rusz et~al.}(2005{\natexlab{a}})\citenamefont{Rusz,
  Turek, and Divis}}]{rtd05}
\bibinfo{author}{\bibfnamefont{J.}~\bibnamefont{Rusz}},
  \bibinfo{author}{\bibfnamefont{I.}~\bibnamefont{Turek}}, \bibnamefont{and}
  \bibinfo{author}{\bibfnamefont{M.}~\bibnamefont{Divis}},
  \bibinfo{journal}{Phys. Rev. B} \textbf{\bibinfo{volume}{71}},
  \bibinfo{pages}{174408} (\bibinfo{year}{2005}{\natexlab{a}}).

\bibitem[{\citenamefont{\c{S}a\c{s}\.{i}o\u{g}lu
  et~al.}(2004)\citenamefont{\c{S}a\c{s}\.{i}o\u{g}lu, Sandratskii, and
  Bruno}}]{ssb04}
\bibinfo{author}{\bibfnamefont{E.}~\bibnamefont{\c{S}a\c{s}\.{i}o\u{g}lu}},
  \bibinfo{author}{\bibfnamefont{L.~M.} \bibnamefont{Sandratskii}},
  \bibnamefont{and} \bibinfo{author}{\bibfnamefont{P.}~\bibnamefont{Bruno}},
  \bibinfo{journal}{Phys. Rev. B} \textbf{\bibinfo{volume}{70}},
  \bibinfo{pages}{024427} (\bibinfo{year}{2004}).

\bibitem[{\citenamefont{Anderson}(1963)}]{Anderson_SolidStatePhysics14}
\bibinfo{author}{\bibfnamefont{P.~W.} \bibnamefont{Anderson}},
  \emph{\bibinfo{title}{Theory of Magnetic Exchange Interactions: Exchange in
  Insulators and Semiconductors}}, vol.~\bibinfo{volume}{14} of
  \emph{\bibinfo{series}{Solid State Physics}} (\bibinfo{publisher}{Academic
  Press, New York}, \bibinfo{year}{1963}).

\bibitem[{\citenamefont{Landau and Binder}(2000)}]{MC-Buch-neu}
\bibinfo{author}{\bibfnamefont{D.}~\bibnamefont{Landau}} \bibnamefont{and}
  \bibinfo{author}{\bibfnamefont{K.}~\bibnamefont{Binder}},
  \emph{\bibinfo{title}{A Guide to Monte Carlo Simulations in Statistical
  Physics}} (\bibinfo{publisher}{Cambridge University Press},
  \bibinfo{year}{2000}).

\bibitem[{\citenamefont{Metropolis et~al.}(1953)\citenamefont{Metropolis,
  Rosenbluth, Rosenbluth, Teller, and Teller}}]{Metropolis}
\bibinfo{author}{\bibfnamefont{N.}~\bibnamefont{Metropolis}},
  \bibinfo{author}{\bibfnamefont{A.}~\bibnamefont{Rosenbluth}},
  \bibinfo{author}{\bibfnamefont{M.}~\bibnamefont{Rosenbluth}},
  \bibinfo{author}{\bibfnamefont{A.}~\bibnamefont{Teller}}, \bibnamefont{and}
  \bibinfo{author}{\bibfnamefont{E.}~\bibnamefont{Teller}},
  \bibinfo{journal}{J. Chem. Phys.} \textbf{\bibinfo{volume}{21}},
  \bibinfo{pages}{1087} (\bibinfo{year}{1953}).

\bibitem[{\citenamefont{Xue et~al.}(1988)\citenamefont{Xue, Grest, Cohen,
  Sinha, and Soukoulis}}]{PhysRevB.38.6868}
\bibinfo{author}{\bibfnamefont{W.}~\bibnamefont{Xue}},
  \bibinfo{author}{\bibfnamefont{G.~S.} \bibnamefont{Grest}},
  \bibinfo{author}{\bibfnamefont{M.~H.} \bibnamefont{Cohen}},
  \bibinfo{author}{\bibfnamefont{S.~K.} \bibnamefont{Sinha}}, \bibnamefont{and}
  \bibinfo{author}{\bibfnamefont{C.}~\bibnamefont{Soukoulis}},
  \bibinfo{journal}{Phys. Rev. B} \textbf{\bibinfo{volume}{38}},
  \bibinfo{pages}{6868} (\bibinfo{year}{1988}).

\bibitem[{\citenamefont{Binder}(1981)}]{Binder-Kum}
\bibinfo{author}{\bibfnamefont{K.}~\bibnamefont{Binder}}, \bibinfo{journal}{Z.
  Phys. E} \textbf{\bibinfo{volume}{43}}, \bibinfo{pages}{361}
  (\bibinfo{year}{1981}).

\bibitem[{\citenamefont{Wan et~al.}(2006)\citenamefont{Wan, Yin, and
  Savrasov}}]{wys06}
\bibinfo{author}{\bibfnamefont{X.}~\bibnamefont{Wan}},
  \bibinfo{author}{\bibfnamefont{Q.}~\bibnamefont{Yin}}, \bibnamefont{and}
  \bibinfo{author}{\bibfnamefont{S.~Y.} \bibnamefont{Savrasov}},
  \bibinfo{journal}{Phys. Rev. Lett.} \textbf{\bibinfo{volume}{97}},
  \bibinfo{pages}{266403} (\bibinfo{year}{2006}).

\bibitem[{\citenamefont{Harrison}(2007)}]{h07}
\bibinfo{author}{\bibfnamefont{W.~A.} \bibnamefont{Harrison}},
  \bibinfo{journal}{Phys. Rev. B} \textbf{\bibinfo{volume}{76}},
  \bibinfo{pages}{054417} (\bibinfo{year}{2007}).

\bibitem[{Qua()}]{QuantumScaling_note}
\bibinfo{note}{For the MC simulations this factor for the Hamiltonian
  corresponds to a scaling of the temperature with the same factor. A quantum
  mechanical calculation for the MFA yields \cite{Anderson_SolidStatePhysics14}
  $ \Theta^{AB}= \frac{2}{3 k_B} \frac{(S+1)}{S} J^{AB}(0)$ for the TMO. So
  going from classical to quantum treatment also corresponds to multiplying
  with $(S+1)/S$. Therefore, we multiplied the classical RPA result with
  $(S+1)/S$, analogous to the MFA. The values obtained from this show excellent
  agreement with such we get by using the RPA approach by Lines\cite{l64},
  which is a quantum approach that, however, considers only nearest and
  next-nearest neighbour interaction and is valid only for the TMO and
  materials with the same magnetic structure.}

\bibitem[{\citenamefont{Solovyev and Terakura}(1998)}]{st98}
\bibinfo{author}{\bibfnamefont{I.~V.} \bibnamefont{Solovyev}} \bibnamefont{and}
  \bibinfo{author}{\bibfnamefont{K.}~\bibnamefont{Terakura}},
  \bibinfo{journal}{Phys. Rev. B} \textbf{\bibinfo{volume}{58}},
  \bibinfo{pages}{15496} (\bibinfo{year}{1998}).

\bibitem[{\citenamefont{Landoldt-B\"ornstein}(1992)}]{landolt-b.III.27g}
\bibinfo{author}{\bibfnamefont{G.~I.} \bibnamefont{Landoldt-B\"ornstein},
  \bibfnamefont{New~Series}}, \emph{\bibinfo{title}{Numerical Data and
  Functional Relations in Science and Technology}}, vol. \bibinfo{volume}{27g,
  Various Other Oxides} (\bibinfo{publisher}{Springer Verlag},
  \bibinfo{year}{1992}).

\bibitem[{\citenamefont{Jauch and Reehuis}(2003)}]{PhysRevB.67.184420}
\bibinfo{author}{\bibfnamefont{W.}~\bibnamefont{Jauch}} \bibnamefont{and}
  \bibinfo{author}{\bibfnamefont{M.}~\bibnamefont{Reehuis}},
  \bibinfo{journal}{Phys. Rev. B} \textbf{\bibinfo{volume}{67}},
  \bibinfo{pages}{184420} (\bibinfo{year}{2003}).

\bibitem[{\citenamefont{Fjellvag et~al.}(1996)\citenamefont{Fjellvag, Gronvold,
  Stolen, and Hauback}}]{fgs96}
\bibinfo{author}{\bibfnamefont{H.}~\bibnamefont{Fjellvag}},
  \bibinfo{author}{\bibfnamefont{F.}~\bibnamefont{Gronvold}},
  \bibinfo{author}{\bibfnamefont{S.}~\bibnamefont{Stolen}}, \bibnamefont{and}
  \bibinfo{author}{\bibfnamefont{B.}~\bibnamefont{Hauback}},
  \bibinfo{journal}{Journal of Solid State Chemistry}
  \textbf{\bibinfo{volume}{124}}, \bibinfo{pages}{52} (\bibinfo{year}{1996}).

\bibitem[{\citenamefont{Roth}(1958)}]{PhysRev.110.1333}
\bibinfo{author}{\bibfnamefont{W.~L.} \bibnamefont{Roth}},
  \bibinfo{journal}{Phys. Rev.} \textbf{\bibinfo{volume}{110}},
  \bibinfo{pages}{1333} (\bibinfo{year}{1958}).

\bibitem[{\citenamefont{Jauch and Reehuis}(2002)}]{jr02}
\bibinfo{author}{\bibfnamefont{W.}~\bibnamefont{Jauch}} \bibnamefont{and}
  \bibinfo{author}{\bibfnamefont{M.}~\bibnamefont{Reehuis}},
  \bibinfo{journal}{Phys. Rev. B} \textbf{\bibinfo{volume}{65}},
  \bibinfo{pages}{125111} (\bibinfo{year}{2002}).

\bibitem[{\citenamefont{Cheetham and Hope}(1983)}]{PhysRevB.27.6964}
\bibinfo{author}{\bibfnamefont{A.~K.} \bibnamefont{Cheetham}} \bibnamefont{and}
  \bibinfo{author}{\bibfnamefont{D.~A.~O.} \bibnamefont{Hope}},
  \bibinfo{journal}{Phys. Rev. B} \textbf{\bibinfo{volume}{27}},
  \bibinfo{pages}{6964} (\bibinfo{year}{1983}).

\bibitem[{\citenamefont{Fang et~al.}(1999)\citenamefont{Fang, Solovyev, Sawada,
  and Terakura}}]{fss99}
\bibinfo{author}{\bibfnamefont{Z.}~\bibnamefont{Fang}},
  \bibinfo{author}{\bibfnamefont{I.~V.} \bibnamefont{Solovyev}},
  \bibinfo{author}{\bibfnamefont{H.}~\bibnamefont{Sawada}}, \bibnamefont{and}
  \bibinfo{author}{\bibfnamefont{K.}~\bibnamefont{Terakura}},
  \bibinfo{journal}{Phys. Rev. B} \textbf{\bibinfo{volume}{59}},
  \bibinfo{pages}{762} (\bibinfo{year}{1999}).

\bibitem[{\citenamefont{Pepy}(1974)}]{p74}
\bibinfo{author}{\bibfnamefont{G.}~\bibnamefont{Pepy}}, \bibinfo{journal}{J.
  Phys. Chem. Sol.} \textbf{\bibinfo{volume}{35}} (\bibinfo{year}{1974}),
  \bibinfo{note}{47}.

\bibitem[{\citenamefont{Kugel et~al.}(1978)\citenamefont{Kugel, Hennion, and
  Carabatos}}]{khc78}
\bibinfo{author}{\bibfnamefont{G.~E.} \bibnamefont{Kugel}},
  \bibinfo{author}{\bibfnamefont{B.}~\bibnamefont{Hennion}}, \bibnamefont{and}
  \bibinfo{author}{\bibfnamefont{C.}~\bibnamefont{Carabatos}},
  \bibinfo{journal}{Phys. Rev. B} \textbf{\bibinfo{volume}{18}},
  \bibinfo{pages}{1317} (\bibinfo{year}{1978}).

\bibitem[{\citenamefont{Tomiyasu and Itoh}(2006)}]{ti06}
\bibinfo{author}{\bibfnamefont{K.}~\bibnamefont{Tomiyasu}} \bibnamefont{and}
  \bibinfo{author}{\bibfnamefont{S.}~\bibnamefont{Itoh}}, \bibinfo{journal}{J.
  Phys. Soc. Jpn.} \textbf{\bibinfo{volume}{75}}, \bibinfo{pages}{084708}
  (\bibinfo{year}{2006}), \bibinfo{note}{43}.

\bibitem[{\citenamefont{Rechtin and Averbach}(1972)}]{PhysRevB.6.4294}
\bibinfo{author}{\bibfnamefont{M.~D.} \bibnamefont{Rechtin}} \bibnamefont{and}
  \bibinfo{author}{\bibfnamefont{B.~L.} \bibnamefont{Averbach}},
  \bibinfo{journal}{Phys. Rev. B} \textbf{\bibinfo{volume}{6}},
  \bibinfo{pages}{4294} (\bibinfo{year}{1972}).

\bibitem[{\citenamefont{Shanker and Singh}(1973)}]{PhysRevB.7.5000}
\bibinfo{author}{\bibfnamefont{R.}~\bibnamefont{Shanker}} \bibnamefont{and}
  \bibinfo{author}{\bibfnamefont{R.~A.} \bibnamefont{Singh}},
  \bibinfo{journal}{Phys. Rev. B} \textbf{\bibinfo{volume}{7}},
  \bibinfo{pages}{5000} (\bibinfo{year}{1973}).

\bibitem[{\citenamefont{Hutchings and Samuelson}(1972)}]{Hutchings_Samuelson}
\bibinfo{author}{\bibfnamefont{M.}~\bibnamefont{Hutchings}} \bibnamefont{and}
  \bibinfo{author}{\bibfnamefont{E.}~\bibnamefont{Samuelson}},
  \bibinfo{journal}{Phys. Rev. B} \textbf{\bibinfo{volume}{6}},
  \bibinfo{pages}{3447} (\bibinfo{year}{1972}).

\bibitem[{\citenamefont{Moreira et~al.}(2002)\citenamefont{Moreira, Illas, and
  Martin}}]{PhysRevB.65.155102}
\bibinfo{author}{\bibfnamefont{I.~P.~R.} \bibnamefont{Moreira}},
  \bibinfo{author}{\bibfnamefont{F.}~\bibnamefont{Illas}}, \bibnamefont{and}
  \bibinfo{author}{\bibfnamefont{R.~L.} \bibnamefont{Martin}},
  \bibinfo{journal}{Phys. Rev. B} \textbf{\bibinfo{volume}{65}},
  \bibinfo{pages}{155102} (\bibinfo{year}{2002}).

\bibitem[{\citenamefont{Shallcross et~al.}(2005)\citenamefont{Shallcross,
  Kissavos, Meded, and Ruban}}]{PhysRevB.72.104437}
\bibinfo{author}{\bibfnamefont{S.}~\bibnamefont{Shallcross}},
  \bibinfo{author}{\bibfnamefont{A.~E.} \bibnamefont{Kissavos}},
  \bibinfo{author}{\bibfnamefont{V.}~\bibnamefont{Meded}}, \bibnamefont{and}
  \bibinfo{author}{\bibfnamefont{A.~V.} \bibnamefont{Ruban}},
  \bibinfo{journal}{Phys. Rev. B} \textbf{\bibinfo{volume}{72}},
  \bibinfo{pages}{104437} (\bibinfo{year}{2005}).

\bibitem[{\citenamefont{Goodenough}(1963)}]{Goodenough_Magnetism_book}
\bibinfo{author}{\bibfnamefont{J.~B.} \bibnamefont{Goodenough}},
  \emph{\bibinfo{title}{Magnetism and the Chemical Bond}}
  (\bibinfo{publisher}{Interscience, New York}, \bibinfo{year}{1963}).

\bibitem[{\citenamefont{Seehra and Giebultowicz}(1988)}]{sg88}
\bibinfo{author}{\bibfnamefont{M.~S.} \bibnamefont{Seehra}} \bibnamefont{and}
  \bibinfo{author}{\bibfnamefont{T.~M.} \bibnamefont{Giebultowicz}},
  \bibinfo{journal}{Phys. Rev. B} \textbf{\bibinfo{volume}{38}},
  \bibinfo{pages}{11898} (\bibinfo{year}{1988}).

\bibitem[{\citenamefont{Rusz et~al.}(2005{\natexlab{b}})\citenamefont{Rusz,
  Turek, and Divi\v{s}}}]{Tkrit-RPA-Rusz}
\bibinfo{author}{\bibfnamefont{J.}~\bibnamefont{Rusz}},
  \bibinfo{author}{\bibfnamefont{I.}~\bibnamefont{Turek}}, \bibnamefont{and}
  \bibinfo{author}{\bibfnamefont{M.}~\bibnamefont{Divi\v{s}}},
  \bibinfo{journal}{Phys. Rev. B.} \textbf{\bibinfo{volume}{71}},
  \bibinfo{pages}{174408} (\bibinfo{year}{2005}{\natexlab{b}}).

\bibitem[{\citenamefont{Towler et~al.}(1994)\citenamefont{Towler, Allan,
  Harrison, Saunders, Mackrodt, and Apra}}]{tah94}
\bibinfo{author}{\bibfnamefont{M.~D.} \bibnamefont{Towler}},
  \bibinfo{author}{\bibfnamefont{N.~L.} \bibnamefont{Allan}},
  \bibinfo{author}{\bibfnamefont{N.~M.} \bibnamefont{Harrison}},
  \bibinfo{author}{\bibfnamefont{V.~R.} \bibnamefont{Saunders}},
  \bibinfo{author}{\bibfnamefont{W.~C.} \bibnamefont{Mackrodt}},
  \bibnamefont{and} \bibinfo{author}{\bibfnamefont{E.}~\bibnamefont{Apra}},
  \bibinfo{journal}{Phys. Rev. B} \textbf{\bibinfo{volume}{50}},
  \bibinfo{pages}{5041} (\bibinfo{year}{1994}).

\bibitem[{\citenamefont{Murnaghan}(1944)}]{Murnaghan}
\bibinfo{author}{\bibfnamefont{F.~D.} \bibnamefont{Murnaghan}},
  \bibinfo{journal}{Proc Natl Acad Sci U S A} \textbf{\bibinfo{volume}{30}},
  \bibinfo{pages}{244} (\bibinfo{year}{1944}).

\bibitem[{\citenamefont{Ding et~al.}(2006)\citenamefont{Ding, Ren, Chow, Zhang,
  Vogel, Winkler, Xu, Zhao, and Mao}}]{ding:144101}
\bibinfo{author}{\bibfnamefont{Y.}~\bibnamefont{Ding}},
  \bibinfo{author}{\bibfnamefont{Y.}~\bibnamefont{Ren}},
  \bibinfo{author}{\bibfnamefont{P.}~\bibnamefont{Chow}},
  \bibinfo{author}{\bibfnamefont{J.}~\bibnamefont{Zhang}},
  \bibinfo{author}{\bibfnamefont{S.~C.} \bibnamefont{Vogel}},
  \bibinfo{author}{\bibfnamefont{B.}~\bibnamefont{Winkler}},
  \bibinfo{author}{\bibfnamefont{J.}~\bibnamefont{Xu}},
  \bibinfo{author}{\bibfnamefont{Y.}~\bibnamefont{Zhao}}, \bibnamefont{and}
  \bibinfo{author}{\bibfnamefont{H.~K.} \bibnamefont{Mao}},
  \bibinfo{journal}{Phys. Rev. B} \textbf{\bibinfo{volume}{74}},
  \bibinfo{eid}{144101} (pages~\bibinfo{numpages}{4}) (\bibinfo{year}{2006}).

\bibitem[{\citenamefont{Holzapfel and Drickamer}(1969)}]{PhysRev.184.323}
\bibinfo{author}{\bibfnamefont{W.~B.} \bibnamefont{Holzapfel}}
  \bibnamefont{and} \bibinfo{author}{\bibfnamefont{H.~G.}
  \bibnamefont{Drickamer}}, \bibinfo{journal}{Phys. Rev.}
  \textbf{\bibinfo{volume}{184}}, \bibinfo{pages}{323} (\bibinfo{year}{1969}).

\bibitem[{\citenamefont{Bloch et~al.}(1966)\citenamefont{Bloch, Chaisse, and
  Pauthenet}}]{bloch:1401}
\bibinfo{author}{\bibfnamefont{D.}~\bibnamefont{Bloch}},
  \bibinfo{author}{\bibfnamefont{F.}~\bibnamefont{Chaisse}}, \bibnamefont{and}
  \bibinfo{author}{\bibfnamefont{R.}~\bibnamefont{Pauthenet}},
  \bibinfo{journal}{Journal of Applied Physics} \textbf{\bibinfo{volume}{37}},
  \bibinfo{pages}{1401} (\bibinfo{year}{1966}).

\bibitem[{\citenamefont{Sidorov}(1998)}]{sidorov:2174}
\bibinfo{author}{\bibfnamefont{V.~A.} \bibnamefont{Sidorov}},
  \bibinfo{journal}{Applied Physics Letters} \textbf{\bibinfo{volume}{72}},
  \bibinfo{pages}{2174} (\bibinfo{year}{1998}).

\bibitem[{\citenamefont{Okamoto et~al.}(1967)\citenamefont{Okamoto, Fujii,
  Hidaka, and Tatsumoto}}]{JPSJ.23.1174}
\bibinfo{author}{\bibfnamefont{T.}~\bibnamefont{Okamoto}},
  \bibinfo{author}{\bibfnamefont{H.}~\bibnamefont{Fujii}},
  \bibinfo{author}{\bibfnamefont{Y.}~\bibnamefont{Hidaka}}, \bibnamefont{and}
  \bibinfo{author}{\bibfnamefont{E.}~\bibnamefont{Tatsumoto}},
  \bibinfo{journal}{Journal of the Physical Society of Japan}
  \textbf{\bibinfo{volume}{23}}, \bibinfo{pages}{1174} (\bibinfo{year}{1967}).

\bibitem[{\citenamefont{Isaak et~al.}(1993)\citenamefont{Isaak, Cohen, Mehl,
  and Singh}}]{PhysRevB.47.7720}
\bibinfo{author}{\bibfnamefont{D.~G.} \bibnamefont{Isaak}},
  \bibinfo{author}{\bibfnamefont{R.~E.} \bibnamefont{Cohen}},
  \bibinfo{author}{\bibfnamefont{M.~J.} \bibnamefont{Mehl}}, \bibnamefont{and}
  \bibinfo{author}{\bibfnamefont{D.~J.} \bibnamefont{Singh}},
  \bibinfo{journal}{Phys. Rev. B} \textbf{\bibinfo{volume}{47}},
  \bibinfo{pages}{7720} (\bibinfo{year}{1993}).

\bibitem[{\citenamefont{Badro et~al.}(1999)\citenamefont{Badro, Struzhkin, Shu,
  Hemley, Mao, Kao, Rueff, and Shen}}]{bss99}
\bibinfo{author}{\bibfnamefont{J.}~\bibnamefont{Badro}},
  \bibinfo{author}{\bibfnamefont{V.~V.} \bibnamefont{Struzhkin}},
  \bibinfo{author}{\bibfnamefont{J.}~\bibnamefont{Shu}},
  \bibinfo{author}{\bibfnamefont{R.~J.} \bibnamefont{Hemley}},
  \bibinfo{author}{\bibfnamefont{H.-k.} \bibnamefont{Mao}},
  \bibinfo{author}{\bibfnamefont{C.-c.} \bibnamefont{Kao}},
  \bibinfo{author}{\bibfnamefont{J.-P.} \bibnamefont{Rueff}}, \bibnamefont{and}
  \bibinfo{author}{\bibfnamefont{G.}~\bibnamefont{Shen}},
  \bibinfo{journal}{Physical Review Letters} \textbf{\bibinfo{volume}{83}},
  \bibinfo{pages}{4101} (\bibinfo{year}{1999}).

\bibitem[{\citenamefont{Lines}(1964)}]{l64}
\bibinfo{author}{\bibfnamefont{M.}~\bibnamefont{Lines}},
  \bibinfo{journal}{Phys. Rev.} \textbf{\bibinfo{volume}{135}},
  \bibinfo{pages}{A1336} (\bibinfo{year}{1964}).

\end{thebibliography}

\end{document}